\documentclass[trackchanges, twocolumn]{aastex701}

\usepackage{graphicx}
\usepackage{amsmath}

\begin{document}

\title{Discovery of the Infrared Counterpart of the Rapid Burster with JWST}

\author[orcid=0009-0003-4448-3681, gname='Malina', sname='Desai']{Malina M. Desai}
\affiliation{Department of Physics, Massachusetts Institute of Technology, Cambridge, MA 02139, USA}
\affiliation{Kavli Institute for Astrophysics and Space Research, Massachusetts Institute of Technology, Cambridge, MA 02139, USA}
\email[show]{mmdesai@mit.edu}  

\author[orcid=0000-0002-7226-836X, gname=Kevin, sname='Burdge']{Kevin B. Burdge} 
\affiliation{Department of Physics, Massachusetts Institute of Technology, Cambridge, MA 02139, USA}
\affiliation{Kavli Institute for Astrophysics and Space Research, Massachusetts Institute of Technology, Cambridge, MA 02139, USA}
\email{kburdge@mit.edu}

\author[orcid=0000-0002-7104-2107, gname='Cristina', sname='Pallanca']{Cristina Pallanca}
\affiliation{Dipartimento di Fisica e Astronomia, Universit\`a di Bologna, Via Gobetti 93/2, Bologna I-40129, Italy}
\affiliation{INAF, Osservatorio di Astrofisica e Scienza dello Spazio di Bologna, Via Gobetti 93/3, 40129 Bologna, Italy}
\email{cristina.pallanca3@unibo.it}

\author[orcid=0000-0003-3182-5569,gname='Saul', sname='Rappaport']{Saul Rappaport} 
\affiliation{Department of Physics, Massachusetts Institute of Technology, Cambridge, MA 02139, USA}
\affiliation{Kavli Institute for Astrophysics and Space Research, Massachusetts Institute of Technology, Cambridge, MA 02139, USA}
\email{sar@mit.edu}

\author[orcid=0000-0002-2165-8528, gname='Francesco', sname='Ferraro']{Francesco R. Ferraro}
\affiliation{Dipartimento di Fisica e Astronomia, Universit\`a di Bologna, Via Gobetti 93/2, Bologna I-40129, Italy}
\affiliation{INAF, Osservatorio di Astrofisica e Scienza dello Spazio di Bologna, Via Gobetti 93/3, 40129 Bologna, Italy}
\email{francesco.ferraro3@unibo.it}

\author[orcid=0000-0001-5613-4938, gname='Barbara', sname='Lanzoni']{Barbara Lanzoni}
\affiliation{Dipartimento di Fisica e Astronomia, Universit\`a di Bologna, Via Gobetti 93/2, Bologna I-40129, Italy}
\affiliation{INAF, Osservatorio di Astrofisica e Scienza dello Spazio di Bologna, Via Gobetti 93/3, 40129 Bologna, Italy}
\email{barbara.lanzoni3@unibo.it}

\author[orcid=0000-0002-0940-6563, gname='Mason', sname='Ng']{Mason Ng}
\affiliation{Department of Physics, McGill University, 3600 rue University, Montréal, QC H3A 2T8, Canada}
\affiliation{Trottier Space Institute, McGill University, 3550 rue University, Montréal, QC H3A 2A7, Canada}
\email{mason.ng@mcgill.ca}

\author[orcid=0000-0002-2218-2306, gname='Paul', sname='Draghis']{Paul Draghis} 
\affiliation{Department of Physics, Massachusetts Institute of Technology, Cambridge, MA 02139, USA}
\affiliation{Kavli Institute for Astrophysics and Space Research, Massachusetts Institute of Technology, Cambridge, MA 02139, USA}
\email{pdraghis@mit.edu}

\author[orcid=0000-0003-4780-4105, gname=Emma, sname=Chickles]{Emma T. Chickles} 
\affiliation{Department of Physics, Massachusetts Institute of Technology, Cambridge, MA 02139, USA}
\affiliation{Kavli Institute for Astrophysics and Space Research, Massachusetts Institute of Technology, Cambridge, MA 02139, USA}
\email{echickle@mit.edu}

\author[orcid=0000-0002-4770-5388, gname=Ilaria, sname='Caiazzo']{Ilaria Caiazzo} 
\affiliation{Institute of Science and Technology Austria (ISTA), Am Campus 1, 3400 Klosterneuburg, Austria}
\email{ilaria.caiazzo@ist.ac.at}

\author[orcid=0000-0001-8804-8946, gname='Deepto', sname='Chakrabarty']{Deepto Chakrabarty}
\affiliation{Department of Physics, Massachusetts Institute of Technology, Cambridge, MA 02139, USA}
\affiliation{Kavli Institute for Astrophysics and Space Research, Massachusetts Institute of Technology, Cambridge, MA 02139, USA}
\email{deepto@mit.edu}

\begin{abstract}

\noindent The Rapid Burster (MXB~1730$-$335) is the only known neutron star to produce both Type I and Type II X-ray bursts, making it a key system for understanding accretion physics onto neutron stars. It resides in Liller 1, a dense, metal-rich stellar system whose crowded core and $\sim$11 magnitudes of visual extinction have concealed the Rapid Burster's optical/infrared (OIR) counterpart for nearly five decades. Here, we present infrared observations of Liller 1 obtained with JWST/NIRCam that, for the first time, reveal the unambiguous OIR counterpart to the Rapid Burster. Caught in an X-ray active state, the counterpart exhibits dramatic infrared variability, fading by $\sim$ 2.5 magnitudes in JWST's F200W band (1.755$-$2.227 $\mu$m) with strong fluctuations on minute timescales. These observations provide an infrared anchor for the Rapid Burster and open the way for future spectroscopy and simultaneous IR and X-ray monitoring to constrain the donor's spectral type and the system's accretion physics. 

\end{abstract}

\keywords{\uat{Compact binary stars}{283} --- \uat{X-ray bursters}{1813} --- \uat{X-ray binary stars}{1811} --- \uat{X-ray sources}{1822} --- \uat{Near infrared astronomy}{1093} --- \uat{High Energy astrophysics}{739} --- \uat{Globular star clusters}{656}}

\section{Introduction} 

Fifty years ago, \cite{1976ApJ...207L..95L} reported the discovery of over 2,000 X-ray bursts within a four day period captured with the Small Astronomy Satellite 3 (SAS$-$3), originating from the neutron star MXB~1730$-$335. Colloquially known as the ``Rapid Burster" (henceforth the RB), this source is the only known neutron star that displays both Type I and Type II X-ray bursts, driven by mass accretion from a companion star (\citealp{1978Natur.271..630H}, \citealp{1976ApJ...210L..13H}). 

As helium (in some cases, hydrogen) accretes onto the surface of a neutron star, the built up layer of material ignites and triggers unstable thermonuclear runaways known as Type I X-ray bursts (\citealp{1977Natur.270..310J}, \citealp{1978ApJ...225L.123J}, \citealp{2007ApJ...661..468C}). This release of energy can occur quasi-periodically (see reviews in \citealp{1993SSRv...62..223L}, \citealp{2021ASSL..461..209G}.) As of October 2025, 122 sources have been cataloged as thermonuclear burst sources (\citealp{2020ApJS..249...32G}, \citealp{intZand2025bursterlist}). Type II X-ray burst observations are significantly rarer, with only two confirmed sources exhibiting them: the RB and the Bursting Pulsar GRO~J1744$-$28 (\citealp{1996IAUC.6286....1K}, \citealp{1996Natur.381..291F}). Type II X-ray bursts originate from accretion flow instabilities that release gravitational energy \citep{1978Natur.271..630H}. 

The RB's Type II bursts display a striking relaxation-oscillation behavior: the fluence (time-integrated flux) is proportional to the waiting time to the next burst, with burst patterns recurring on 1--10~minute cycles during activity intervals lasting $\sim$100--200~days \citep{1993SSRv...62..223L, 2002A&A...381L..45M, 2015MNRAS.449..268B}. GRO~J1744$-$28 is a rapidly rotating neutron star with a 0.467 second spin period in a 11.76 day orbital period with a low-mass companion \citep{1996IAUC.6286....1K}. Its X-ray bursts last around 10 seconds each at a rate of roughly 2 to 20 bursts per hour (\citealp{1997ApJ...486..435R}, \citealp{1996ApJ...469L..29Z}), while the RB can burst up to thousands of times per day \citep{1999MNRAS.307..179G}. While GRO~J1744$-$28 exhibits coherent X-ray bursts phased to its spin period, the RB shows no such periodicity when alternating between active bursting episodes and quiescent intervals. Studies comparing the two sources have suggested that the RB's lack of coherent pulsations may be explained by a weaker magnetic field relative to GRO~J1744$-$28 \citep{1996ApJ...462L..39L}, which would allow accreted material to spread more uniformly over the neutron star surface rather than being channeled onto the magnetic poles. Additionally, the neutron star SMC X$-$1 is theorized to emit bursts that phenomenologically appear as Type II bursts, though the mechanism underlying these explosions is likely different than those present in the Rapid Burster (\citealp{1997A&A...321L..25L}, \citealp{2003ApJ...582L..91M}, \citealp{2018RAA....18..148R}, \citealp{2020ApJ...895...10P}).

Due to its unique nature, the RB has been observed for several decades across many wavelengths. The initial search for the optical companion led to the discovery of Liller 1 \citep{1977ApJ...213L..21L}. Initially cataloged as a globular cluster, this stellar system has been recently discovered to host distinct sub-populations with age differences of $\sim 10$ Gyr \citep{ferraro+21, dalessandro+22}, iron abundances varying from [Fe/H]$\sim -0.48$ to $+0.26$ \citep{crociati+23}, and abundance patterns fully consistent with those of bulge field stars \citep{deimer+24, fanelli+24, ferraro+25, chiappino+26}. For these reasons, it is now considered, together with Terzan 5 \citep{ferraro+09, ferraro+16, origlia+25, zullo+26}, to be a bulge fossil fragment \citep{ferraro+21}, i.e., the remnant of a primordial structure that contributed to the early assembly of the Galactic bulge.

The RB's membership in Liller 1 has made the search for a binary counterpart difficult: Liller 1 is known to be metal-rich, dense, and extremely reddened by Galactic dust \citep{1995AJ....109.1154F}. These features make Liller 1 difficult to observe at optical wavelengths, but encourage such searches in the infrared \citep{2001AJ....122.2627H}. Observations have been performed in the optical  and the near-infrared using Hubble Space Telescope's (HST) instruments NICMOS \citep{2001A&A...376..878O} and ACS \citep{ferraro2019_hst15231, ferraro+21}, and the Gemini South Adaptive Optics Imager (GSAOI) on the Gemini South Telescope, enhanced with the GeMS adaptive optics system \citep{2015ApJ...806..152S}.

Most recently, \cite{Pallanca2025} (hereafter P25) identified a variable source in the vicinity of the RB's X-ray and radio position using HST F606W (0.47–0.70 $\mu$m) and F814W (0.69–0.89 $\mu$m) imaging together with Gemini $K_s$ (1.99–2.31 $\mu$m) data, and proposed it as the most promising near-infrared counterpart candidate to date. Notably, these observations were taken while the RB itself was in quiescence.

Here, we introduce new near- and mid-infrared observations of Liller 1 taken with JWST's NIRCam instrument. Our time-series observations serendipitously captured Liller 1 while the RB was in outburst (see Figure~\ref{fig:maxi}), clearly revealing its highly variable infrared counterpart, a distinct source from the previous candidate reported in P25. We also identify P25's source and find that it is likely an unrelated periodic contact binary system distinct from the RB. Through our variability search described in Section~\ref{ssec:vardet}, we localize and provide photometry of the RB's true counterpart, consistent with previous localization constraints from Chandra \citep{2001AJ....122.2627H, 2024ApJS..274...22E}, and interpret these findings.

\section{Observations}\label{sec:obs}

\begin{figure*}[ht!]
    \centering
    \includegraphics[width = \textwidth]{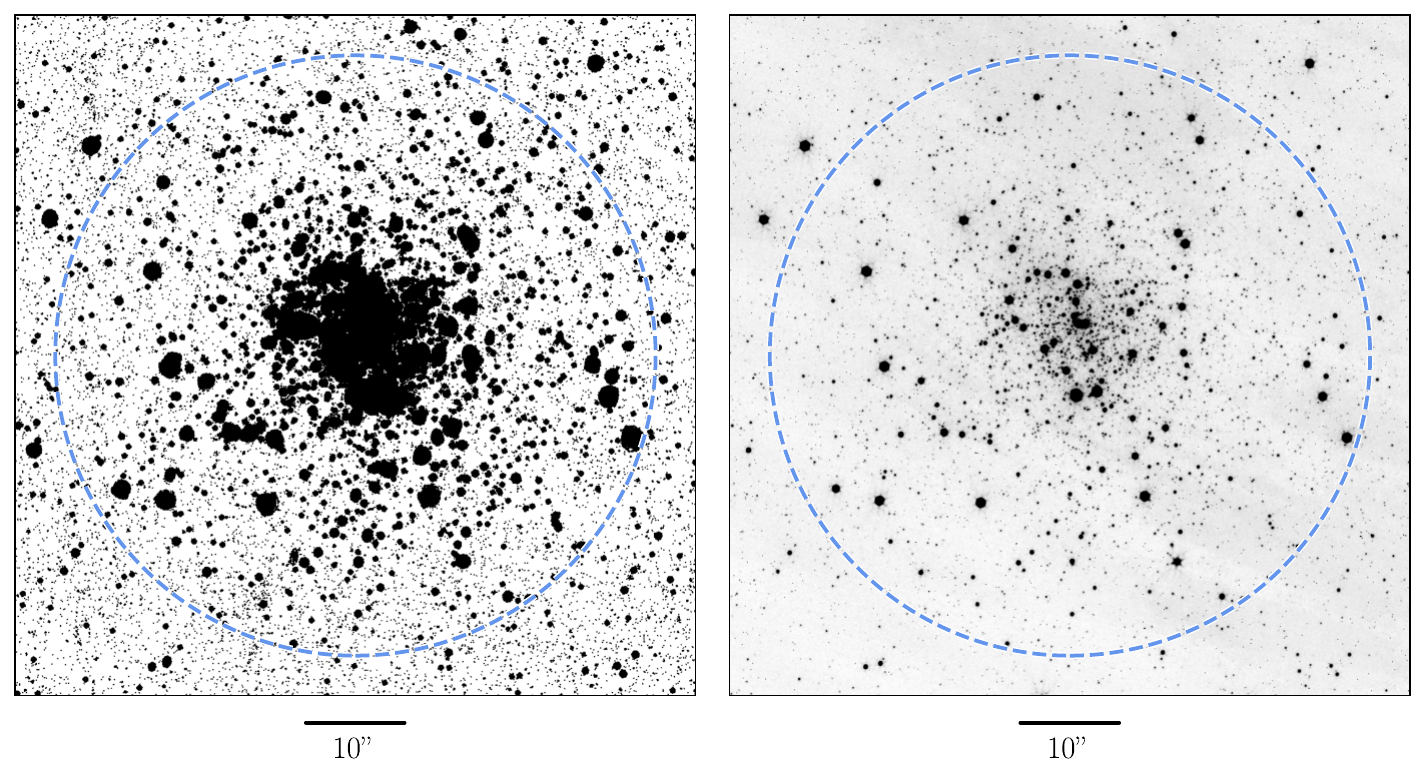}
    \caption{JWST calibrated (.cal) file (left panel) and zeroframe (right panel) data of Liller 1 from JWST's NRCBLONG detector. The half-mass radius of 30.5$''$ has been inscribed for reference, with length scales in arcseconds below each panel. North is up, and East is to the left. We find saturation and crowding issues are minimized when Liller 1 is viewed in the zeroframe image.}
    \label{fig:fov}
\end{figure*}

We analyze data from JWST Proposal 5381 (PI: Kevin Burdge), taken in two segments on 2025 April 22 and 2025 April 23 during an RB outburst (see MAXI data, Figure~\ref{fig:maxi}). Data from Liller 1 were recorded in two undithered visits separated by 27.7 hours with NIRCam in the F200W (short wavelength, 1.755-2.227 $\mu$m, with detectors NRCB1, NRCB2, NRCB3, and NRCB4) and F356W (long wavelength, 3.136-3.981 $\mu$m, with detector NRCBLONG) bands. Each visit is comprised of 12 exposures each with a total effective exposure time of 1932.6 seconds, resulting in a total observation time of 12.88 hours. 

The two panels of Figure~\ref{fig:fov} display the NIRCam observations of Liller 1 in the long wavelength filter. The left image displays the default JWST calibrated file on the full exposure. Zooming into the core region, we see that the dense clustering of stars leads to high levels of saturation. Combined with severe crowding of many IR-bright sources, photometry is challenging in the full exposures. JWST offers a non-destructive readout precisely to combat these issues. The observations were taken using the BRIGHT2 pattern, where two frames are taken per group and are averaged into one readout. Each exposure is split into nine integrations, with each integration split into ten groups, resulting in 90 groups of data per exposure. The counterpart saturates the pixels across one group, rendering these 90 averaged frames unusable. In order to perform precise photometry, we instead use the ``zeroframes'', the very first frames of each integration saved before they are averaged with the second frame to form the initial group. The right panel of Figure~\ref{fig:fov}  shows the reduction in core saturation by using the zeroframe data as opposed to the calibrated data of the same region in the left panel. With a zeroframe exposure time of only 10.37 seconds, many of the previously saturated stars become accessible. 

We process the zeroframes through JWST's stage 1 pipeline  \citep{2022zndo...7487203B, bushouse_2025_17400413}. By default, the pipeline masks saturated pixels; corrects for readout drifts, jumps due to cosmic rays, and non-linearity in NIRCam's gain; and performs dark and superbias subtraction. We slightly modify the pipeline to include flicker (1/f) noise correction, which appears as visible banding in the zeroframe images. These effects greatly impact the background noise and can impede precise photometry if left uncorrected. The full data reduction pipeline is described in \cite{KBurdgeinprep}.

We include the default calibrated images from the previous HST Progam 15231 (PI: Francesco Ferraro) taken on 2019 August 17 in the F606W and F814W bands in our photometric analysis. 

\begin{figure}
    \centering
    \includegraphics[width = \columnwidth]{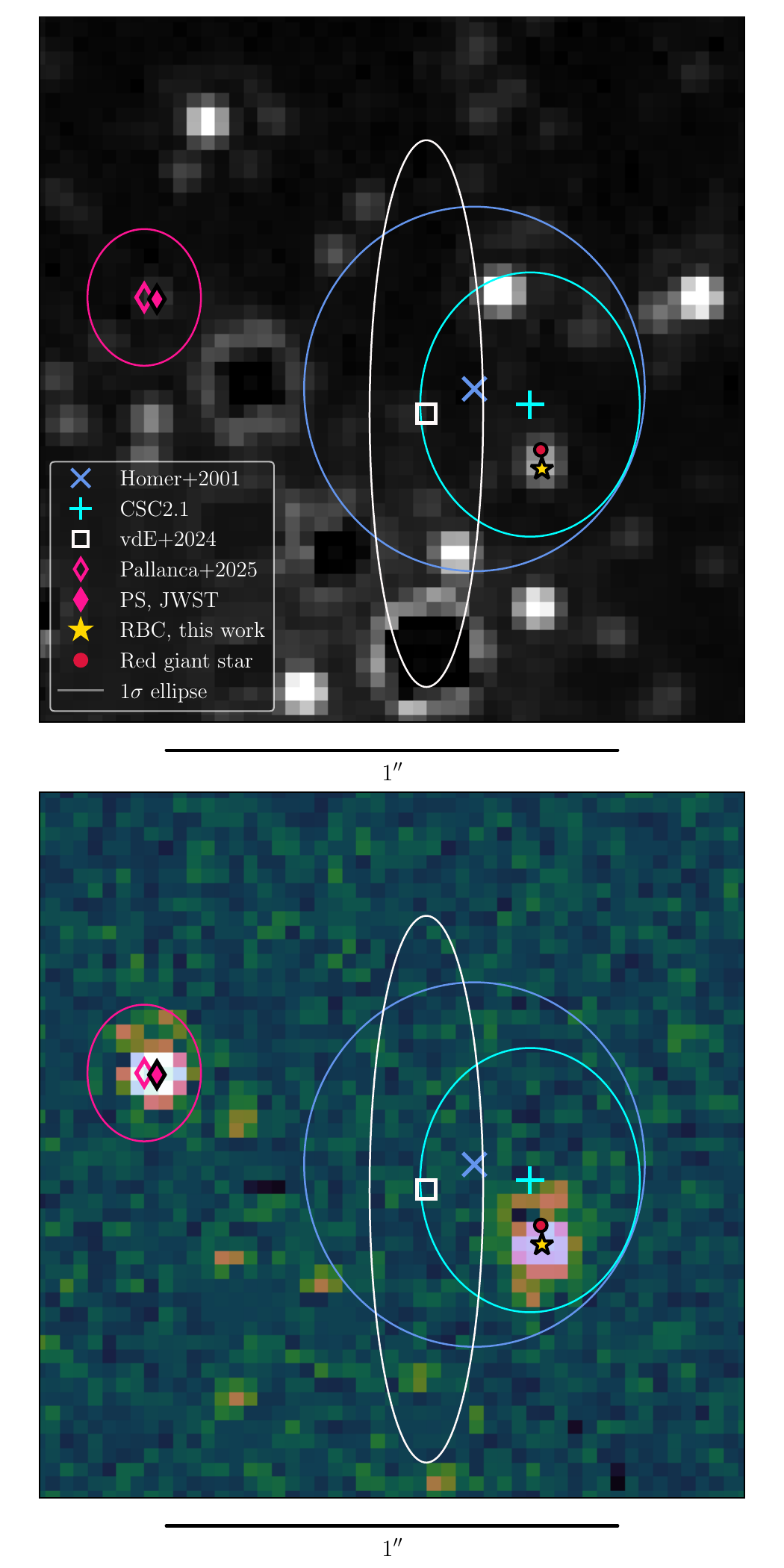}
    \caption{Localization of the Rapid Burster counterpart (labeled RBC, gold star) and a neighboring red giant (red circle). Previous localizations with $1 \sigma$ errors are shown for Chandra (blue; updated CSC2.1 match in cyan; \citealp{2001AJ....122.2627H}) and the VLA (white; \citealp{2024MNRAS.533..756V}). The magenta open diamond marks \cite{Pallanca2025}'s source position with its $1 \sigma$ error ellipse; the filled magenta diamond marks our identification of this source in the JWST data. Top: Noise- and saturation-corrected zeroframe data in the F200W band. Bottom: Peak pixel autocorrelation from the same data, highlighting sources with coherent variability. North is up, east is to the left.}
    \label{fig:localization}
\end{figure}

\section{Methods} \label{sec:phot}

\subsection{Variable Detection}\label{ssec:vardet}

We perform a search for variable stars on the zeroframe data to create an initial catalog of interesting sources. Simple variability detection methods fail due to the non-negligible amount of cosmic rays that JWST observes, which create sharp spikes in flux. To avoid contamination in our catalog from cosmic rays, we calculate the interquartile range (IQR) of each pixel of the median-subtracted zeroframe. The IQR eliminates outliers such as cosmic rays, providing a viable metric of pixel variability. We mask saturated pixels, avoid edge pixels, and require the pixel to be $1\sigma$ above the median background. 

We then compute the autocorrelation of each pixel. Autocorrelation measures the correlation of a single variable and its subsequent value over time. The autocorrelation function (ACF) peaks at time lags where the signal is highly correlated with itself. The ACF is mathematically defined as the expectation value of the mean-subtracted signal $X_{t}$ at some time $t$ multiplied by the complex conjugate of $X_{t + \tau}$, its value after some time $\tau$: 

\begin{equation}
    R_x (\tau) = \langle X_{t} \cdot X_{t+\tau}^* \rangle.
\end{equation}

\noindent Since this value is always real, we can compute the real fast Fourier transform and the inverse real fast Fourier transform to calculate the unnormalized ACF:

\begin{equation}
    \mathcal{F}(\nu)^{-1} {|\mathcal{F}(\nu)|^2}[\tau] = \sum_{t} X_t \cdot X_{t + \tau},  
\end{equation}

\noindent where $\mathcal{F}(\nu)$ is the Fourier transform. An equivalent way of envisioning the ACF is as the inverse Fourier transform of the power spectral density $S(\nu)$:

\begin{equation}
    S(\nu) = |\mathcal{F}(\nu)|^2.
\end{equation}

\noindent We then normalize the ACF by the first value of the function, which is the variance of the pixel.

A high autocorrelation value indicates structured signal, while a low value indicates stochasticity. The autocorrelation for white noise, for example, can be represented by a Dirac delta function at $\tau = 0$. To exclude the pixel's self correlation, we take the series from $\tau = 1$. Thus, the peak ACF of a pixel of this series acts as a proxy for coherent variability. 

We compute the ACF across our 96 zeroframes in each segment and take the peak value for each pixel to create a map of highly variable pixels with structured variability. Our search resulted in a catalog of nearly 900 variable sources, two of which we identify as the counterpart to the RB and P25's source. The bottom panel of Figure~\ref{fig:localization} displays the map of peak ACF in the region of the RB. Both sources show a high peak ACF, while neighboring stars show similar variability to the background pixels. This search method allows for quick visual identification of candidate variable sources.

\subsection{Astrometry Corrections}

To ensure proper localization, we cross-matched our JWST observations with the Gaia DR3 catalog \citep{2023A&A...674A...1G}. We restrict our query with the following cuts:
\begin{itemize}
    \item RUWE $\leq 1.4$  
    \item PM R.A. error $ < 1 $ mas/yr
    \item PM Dec error $ < 1 $ mas/yr
    \item $ 13 < $ G mag $ < 20 $ 
\end{itemize}

Our Renormalized Unit Weight Error (RUWE) cut, informed by \cite{2018RenormalisingTA}, acts as a proxy for selecting single stars by eliminating potential binaries. After these cuts, 812 Gaia sources with proper motion solutions remain in the queried field, of which 299 fall on NRCBLONG and $\sim 70$ on each of the short-wavelength detectors. Gaia DR3 positions (epoch 2016.0) were propagated to the JWST observation epoch ($\Delta t = 9.31$ yr) using per-star proper motions with the standard $\cos\delta$ correction applied to right ascension. For each detector and segment, we fit each match with the appropriate PSF model on a deep stack of 96 zeroframes, initializing each fit at the position projected through the original pipeline WCS. Matches with saturated cores, low signal-to-noise, position excursions $ > 1.5$ px, or poor fits are excluded from the WCS correction, leaving around 40 well-matched sources per short-wavelength detector and 172 for NRCBLONG with a per-star scatter of 4-5 mas. The sigma-clipped median of the difference between the fitted and projected positions provides the WCS shift: for NRCB4, we find a $(\Delta x, \Delta y)$ shift of $(-10, -9)$ and $(-10, -11) \pm 1$ mas for the first and second segments, and a  $(-12, -16)$ and $(-13, -16) \pm 1$ mas shift for NRCBLONG.

\subsection{PSF Photometry}{\label{sec:psf}}

\begin{figure*}
    \centering
    \includegraphics[width = \textwidth]{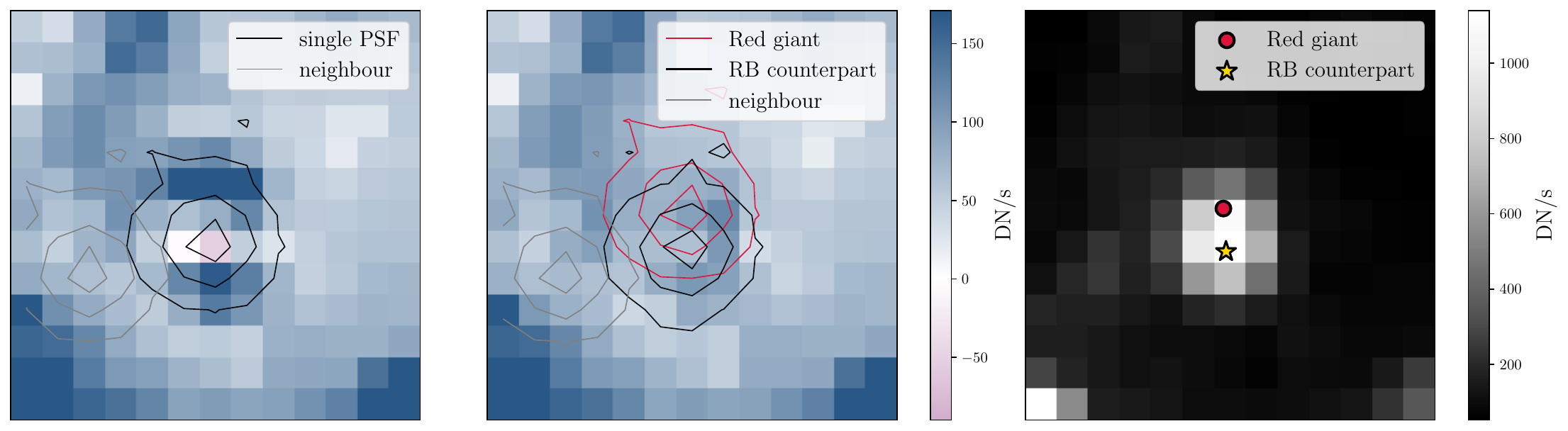}
        \caption{Residuals from a single-source PSF fit (left panel), residuals from a 2-source PSF fit (central panel), NIRCam position of the RB counterpart and the non-variable blended star that we classify as a red giant (right panel). We perform our PSF routine on the 13x13 pixel box shown, with each pixel covering 0.031$''$. Included in gray is a nearby neighboring star whose PSF infringes on the blend. There is a noticeable improvement in the reduction of residual wings in the center of the image when a two-source blend is adopted. North is up, east is left.}
    \label{fig:psffit}
\end{figure*}

\begin{table*}
\centering
\caption{Positions of the Rapid Burster counterpart and comparison sources.
Coordinates are given at each source's own epoch; the PM column is the shift to
epoch 2025.31 using the Liller~1 proper motion
($\mu_{\alpha*},\mu_\delta$)~$=(-5.88,-7.56)$~mas~yr$^{-1}$, and the final column
is the resulting offset from the JWST RB counterpart.}
\label{tab:rb_positions}
\begin{tabular}{llcccccc}
\hline\hline
 & \multicolumn{2}{c}{Position (ICRS)} & \multicolumn{2}{c}{Uncertainty (mas)} & & PM shift \\
Source & RA & Dec & $\sigma_{\alpha*}$ & $\sigma_\delta$ & Epoch & (mas) \\
\hline
Homer et al.\ \citeyear{2001AJ....122.2627H}           & 17h33m24.5993s & $-$33d23m20.084s & 375.7 & 400 & 2000.59 & $(-134,-184)$ \\
CSC2.1 \citep{2024ApJS..274...22E}                     & 17h33m24.5895s & $-$33d23m20.118s & 242.1 & 290 & 2024.25 & $(-6,-8)$     \\
van den Eijnden et al.\ \citeyear{2024MNRAS.533..756V} & 17h33m24.6078s & $-$33d23m20.138s & 125.2 & 600 & 2020.21 & $(-28,-38)$   \\
Pallanca et al.\ \citeyear{Pallanca2025} (P25)         & 17h33m24.6575s & $-$33d23m19.883s & 125.2 & 150 & 2019.50 & $(-31,-43)$   \\
\hline
RB counterpart, this work     & 17h33m24.5874s & $-$33d23m20.259s & 7.0 & 7 & 2025.31 & -- \\
Red giant (A), this work      & 17h33m24.5877s & $-$33d23m20.217s & 7.0 & 7 & 2025.31 & -- \\
P25 source, JWST (this work)  & 17h33m24.6553s & $-$33d23m19.887s & 7.0 & 7 & 2025.31 & -- \\
\hline
\end{tabular}
\end{table*}

DOLPHOT is a photometry tool designed for crowded fields, making it suitable for Liller 1. To robustly perform photometry on Liller 1, we create an initial catalog of sources using the DOLPHOT 3.0 photometric package \citep{2016ascl.soft08013D}, then perform forced photometry where necessary.  The user provides a tuned list of photometric parameters, science images, and a deep reference calibration image to DOLPHOT. The program astrometrically aligns all science images to the reference, performs iterative Point-Spread Function (PSF) photometry, and returns a list of stars with their photometry, flags, and useful fitting statistics. We make use of the NIRCam and ACS modules which contain PSF models for the F200W, F356W, F606W, and F814W bands. Our noise, photometric, and fitting parameters are informed by the results discussed in \cite{2024ApJS..271...47W}, which recommends distinct input parameters for separate channels. 

The resulting catalog of stars is further filtered with the following criteria:

\begin{itemize}
    \item SNR$\geq 4$ 
    \item Sharpness$^2$ $\leq 0.01$
    \item Crowding $\leq 2.25$
    \item Flag $\leq 3$
    \item Type $\leq 2$
\end{itemize}

\noindent where object type is defined by DOLPHOT as: 1 - bright star, 2 - faint star, 3 - elongated source, 4 - narrow source, 5 - extended source. This selection aims to reject outliers while maximizing completeness \citep{2024ApJS..271...47W}. We recover P25's source in all four bands through our catalog. 

While the DOLPHOT catalog provides some coverage of the central region of Liller 1, saturation prevents recovery of stars within its core, and crowding issues can result in blended sources appearing as a single star. This is the case for the RB counterpart centroid. When a single PSF is fit to the source centroid, residual counts left in the wings of the PSF contour are visible, shown in the first panel of Figure~\ref{fig:psffit}. We find that adopting two point sources results in a large reduction of the residual light, which indicates that there is another source blended with the RB counterpart. 

Using the \texttt{stpsf} package, we create a gridded PSF for the F200W filter \citep{2012SPIE.8442E..3DP}. To calibrate our photometry, we perform PSF photometry on a carefully curated sample of 25 reference core cluster stars per visit. These stars must be bright, isolated, and close to the blend. We search for stars 25 times above the standard deviation using \texttt{DAOStarFinder}, limiting the candidates to a 6.2 arcsecond box centered on the blend \citep{1987PASP...99..191S}. These stars must be at least 7 pixels from another detection and the surrounding 11 by 11 pixel box must be composed of clean pixels. To ensure no blends are used for calibration, the PSF rms residual must be less than 0.04. The top 25 brightest stars from each visit are selected and vetted by eye. We use these stars to determine if any frame-to-frame positional shifts occur as the PSF fit to the blend is highly dependent on positional accuracy. 

To fit the blended PSF, we rely on two key pieces of information: the RB was in quiescence during the HST observations and therefore very faint, and the RB is highly variable and visible during the JWST observations. We identify a non-variable source slightly offset from the location of our JWST blended centroid in the HST images, allowing us to pin the source to the north of the RB counterpart. 

We use \texttt{photutils} PSFPhotometry routine \citep{2016ascl.soft09011B} to fit an initial positional guess on a deep stack of each segment's zeroframes. Using 25 nearby reference stars, we calibrate each frame's offset from the median pointing, finding a range of per-frame shifts that have a standard deviation in (x, y) of $(9.1, 6.1)$ mpx in visit 1 and $(9.0, 6.3)$ mpx in visit 2. We correct frame shifts prior to minimizing $\chi^2$ across all frames to find the positions of the blend components.

We also take into account the crowding in Liller 1's core by modelling neighbouring stars. One star, whose PSF footprint is outlined in Figure~\ref{fig:psffit} in gray, is close enough to the blend that it is used in our primary simultaneous PSF fit with the blend itself. We search for neighbouring stars within an 81 x 81 px box centered on the blend in the deep image with \texttt{DAOStarFinder} with a threshold of $8\sigma$. From this search, we model five stars within a 22 px region surrounding the blend that contribute a median of 8.3 DN/s and a max of 675.7 DN/s to pixels within the primary blend fit box. Their flux is subtracted prior to fitting the flux amplitudes of the blend.

Zeroframes face an additional issue that is solved in the calibrated frames: cosmic rays (CRs). The JWST pipeline is able to correct for CRs by using jump detection, which measures the difference in flux between consecutive readouts to select spurious outliers. Because our zeroframes are the first read of each integration, separated by a full integration (225.5 s), jump detection is unavailable and CRs are identified manually. We only check for CR hits within the 13 x 13 primary fit box around the blend's center. By calculating the difference between each pixel's data and modelled flux over the pixel's standard deviation, we can identify unusually large fluctuations for that particular pixel. We subtract each pixel's own median pull across all frames and then take the median absolute deviation (MAD) of each pixel, keeping those that are within a $10\sigma$ outlier threshold to the data. We identify and visually confirm a CR that lands on one of the blend's central pixels in the 33rd frame. Since the RB counterpart is a highly variable source, the frame containing the CR is dropped from the PSF fit to avoid contaminating the resulting light curves so we can report flares confidently.

We fit the blend twice. A first fit minimizes $\chi^2$ with the separation free, but with $\chi^2/\text{DoF}  \approx 130$ it is dominated by mismatch between the model and the true PSF, which biases the separation to smaller values (1.301 and 1.267 px in the two visits) and inflates the flux assigned to the RB counterpart. We therefore fix the separation, which is astrophysically constant, through two criteria. First, the non-variable star's flux must be uncorrelated with the counterpart's within a visit. To determine the separation where this occurs, we displace the counterpart's $x$-position by a trial amount, refit every frame by weighted least squares, and regress the non-variable star's light curve against the counterpart's. The slope of the regression measures the component of the non-variable star's variation proportional to the counterpart's. Second, we expect the non-variable star to have a flat flux consistent between visits. The two criteria do not coincide. The first gives $1.341 \pm 0.011$ px and $1.329 \pm 0.004$ px in the two visits; the second gives 1.357 px. We adopt the latter, since 34 isolated field stars in the same region agree between visits to $1.0002 \pm 0.0003$ indicating that there is no measurable flux difference between the two visits for non-variable stars. At this separation, the non-variable star measures 4212 and 4223 DN/s in the two visits, agreeing to $0.26\%$. We attribute the inconsistency between the two measurements as a flux-dependent response rather than to geometry. 

Multiple sources of errors from the PSF fitting routine are considered and split into two categories: frame-by-frame or relative errors, and systematic errors. The per-pixel uncertainty is the measured readnoise of the zeroframe added in quadrature with the Poisson noise from the total photon count in that pixel. We leave these errors unscaled as they are consistent with white noise measured from the non-variable star. The systematic uncertainty is set by sensitivity to the field model rather than by the statistical error: repeating the fit without the neighbouring stars modelled shifts the separation by 0.047 and 0.042 px in the two visits, and we adopt $\pm0.045$ px as the $1\sigma$ systematic. This dominates both the 0.004$-$0.011 px uncertainty on the separation solution itself and the 0.014$-$0.015 px $\chi^2$ covariance error. While we minimize these errors with the separation fit described above, this contributes a $\pm 5.6 \%$ and $\pm 2.1 \%$ error for the non-variable star and a $\pm2.2\%$ and $\pm2.6\%$ error for the RB counterpart for visit 1 and visit 2, changing based on each visit's brightness contrast between the two objects.

As seen in Figure~\ref{fig:psffit}, we are able to disentangle the blended source and classify one of the components as a red giant star using the HST and Gemini data.  We note that the autocorrelation shown in panel 2 of Figure~\ref{fig:localization} peaks towards the south pole of the blend, consistent with our PSF deblending. Using our calibrated WCS coordinates, we report the localizations of the red giant star and the RB counterpart in Table~\ref{tab:rb_positions}. We also perform an approximate proper motion correction for each of the previous localizations provided in the table by using Liller 1's bulk motion of $(\mu_\alpha, \mu_\delta) = (-5.403, -7.431)$ mas/yr \citep{2021MNRAS.505.5978V}, and use these values to compare localizations in  Figure~\ref{fig:localization}.

A similar fit in the F356W band is more challenging. The corresponding on-sky separation of the two stars is $\sim 42$ mas, well below the F200W FWHM of 2.141 px ($66$ mas). NRCBLONG's pixel resolution of 63 mas/px means that one pixel contains the entirety of the blend. With a coarser sampling and a wider PSF, a two-source fit is degenerate with the blended centroid in the long-wavelength channel and the sources cannot be fully disentangled. Thus, we are unable to provide an infrared F200W-F356W color for the RB counterpart. 

\begin{figure*}[ht!]
    \centering
    \includegraphics[width = \textwidth]{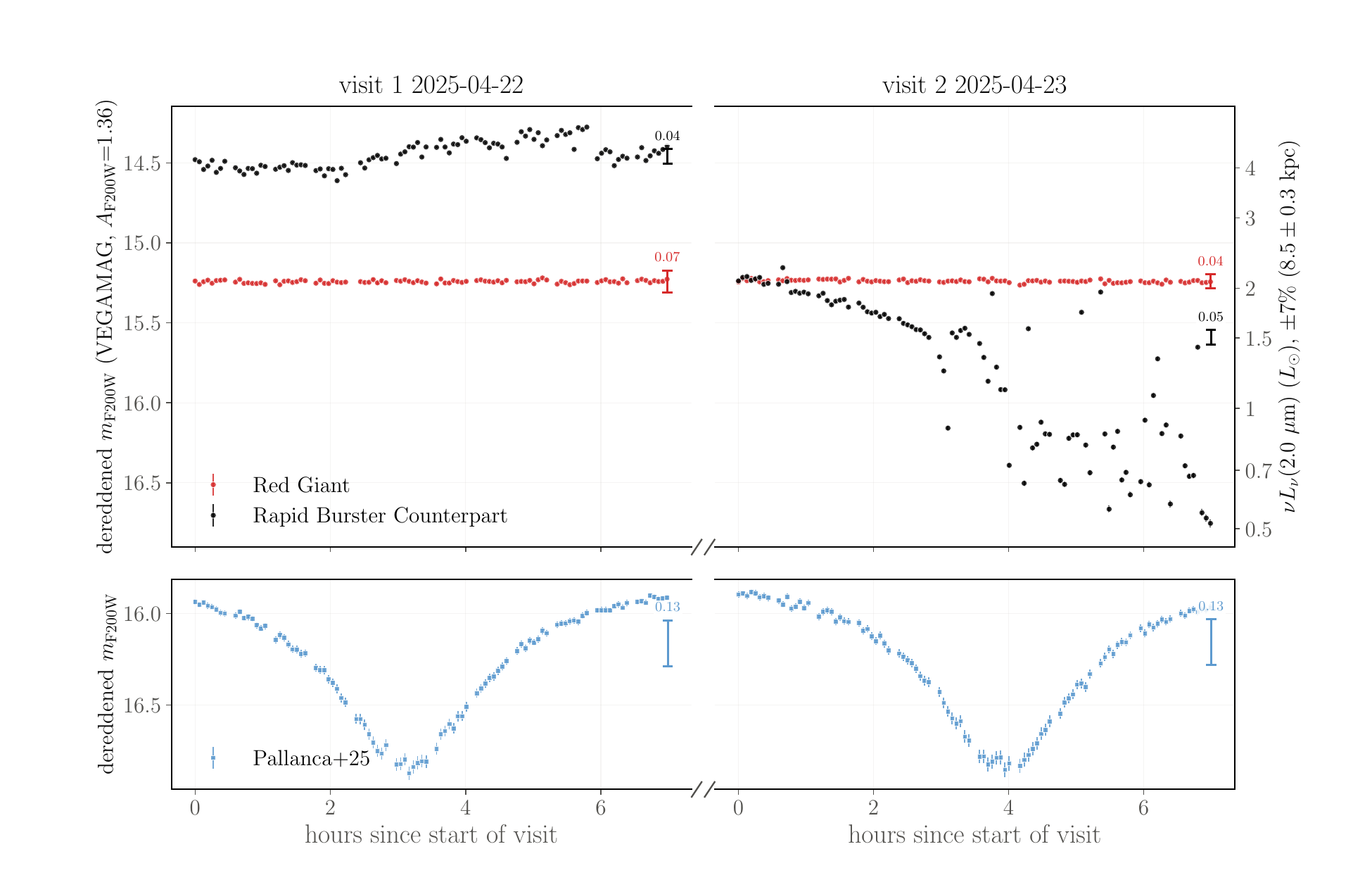}
    \caption{Top panel: light curves for the red giant (in red) and the RB counterpart (in black). Bottom panel: light curve for \cite{Pallanca2025}'s previously identified candidate. Relative frame-to-frame error bars are shown for all points, with systematic error bars shown for each visit per source.}
    \label{fig:lcs}
\end{figure*}

\subsection{Light Curves}\label{sec:lcs}

We extract light curves from the zeroframes using the PSF photometry methods described above. The zeroframe of the first integration of each exposure sits measurably higher than the other zeroframes, so we elect to drop the first zeroframe of the first integration in each exposure. Zeroframes are spaced 225.5s apart, and our data have an inter-exposure gap of 332.8s. As a result, our light curves have 11 gaps of $\sim 560$s between our 12 exposures.

Because the light curves are measured on zeroframe data rather than the pipeline cal products, we must verify if the JWST zeropoint flux calibration transfers without modification. Using the same PSF model as the science photometry, we measure total PSF fluxes for 34 (visit 1) and 33 (visit 2) isolated, unsaturated stars within a box of half-size 150 px centered on the RB counterpart on the visit's deep zeroframe stack and mean cal image. The resulting zeroframe-to-cal flux ratio is $1.0004 \pm 0.0048$, leading us to adopt the calibrated VEGAMAG zeropoint of $25.661 \pm 0.005$ mags for the NRCB4 detector. We note that this is a local calibration specific to the region in Liller 1's core.

Liller 1 is known to be highly extincted. To correct for this, we use a color excess value of $E(B - V) = 4.52 \pm 0.10$ and extinction ratio $R_{V} = 2.5$ as determined by \cite{2021ApJ...917...92P}, resulting in a V-band extinction of 11.3 mags. We adopt \cite{1989ApJ...345..245C}'s extinction law to find a total extinction $A_{\text{F200W}}$ of 1.36 mags at the F200W pivot wavelength of $1.9875 \ \mu$m. We note that adopting the extinction law of \cite{1999PASP..111...63F} instead would give $A_{\text{F200W}} = 1.44$ mag, higher by 0.08 mag ($\sim6\%)$. We adopt Cardelli et al. throughout and do not include this difference in our systematic uncertainties since it shifts all dereddened magnitudes together and cancels in colours and in the variability amplitude. 

The extinction should be weighted across the F200W bandpass, which requires knowing the effective temperature $(T_{\text{eff}})$ of the sources. We evaluate the extinction for spectral energy distributions (SEDs) spanning blackbodies from 3000 to 30000 K and power laws from $f_\nu \propto \nu^{1/3}$ to $f_\nu \propto \nu^{2}$ and find $A_{\text{F200W}} = 1.366$--$1.387$ mag. The pivot value of 1.358 mag underestimates this by 0.019 mag on average, which we carry as a systematic error common to all sources. In this way, we avoid any assumptions on $T_{\text{eff}}$. We propagate our earlier frame-to-frame errors for both sources independently, which are reflected in each point's individual error bars. 

The final Vega magnitude F200W light curves are provided in Figure~\ref{fig:lcs}. The RB counterpart displays remarkable flaring behavior as it suddenly fades by $\sim$2.5 mags during the second visit. Taking the distance to Liller 1 as $8.5 \pm 0.3$ kpc  \citep{ferraro+21}, we calculate the in-band luminosity and find that the change in magnitude corresponds to a decrease of 3.5 $L_\odot$. The variability is present in both the F200W and F356W bands (see the Appendix for further discussion on temporal coherence of the RB counterpart flares). 

We also note that the timescale of flaring is consistent with the decay timescale of Type I bursts. Across visit 2, 6 bright flares are present in the last 4 hours of the observation. Type II bursts have durations ranging from 2.4 to 23.73 s and occur in intervals ranging from 14.4 to 210 s at the 5 to 95 percentile of their distribution \citep{2015MNRAS.449..268B}. As the bulk of Type II bursts have recurrence times $<200$s, our time resolution cannot sufficiently resolve them. Type I bursts, on the other hand, occur at intervals of $\sim$1 hour or occasionally more, and can last over 200s \citep{1993SSRv...62..223L, 1999MNRAS.307..179G, 2017arXiv170307221I}. This timescale aligns with the pattern we observe, though higher time resolution and simultaneous X-ray observations are required to confirm this hypothesis. The aperiodic flaring and rapid decline in brightness over short timescales strongly suggest that this source is the counterpart to the RB.

\subsection{Color-Magnitude Diagram}

\begin{figure}[ht]
    \centering
    \includegraphics[width = \columnwidth]{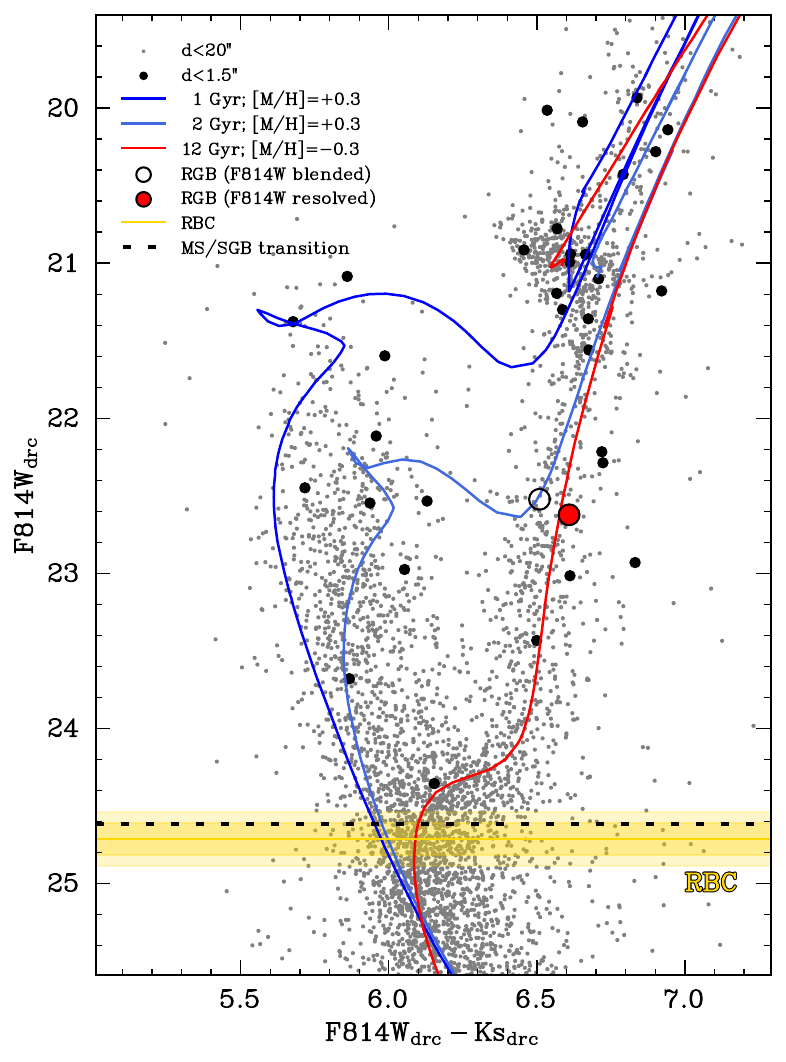}
    \caption{Color-magnitude diagram of Liller 1 from HST's F814W and Gemini's $K_s$ band. Solid curves show 1, 2, and 12 Gyr isochrones representatives of Liller 1's young and old populations. The yellow solid line denotes the F814W magnitude constraint of the RBC in quiescence. The yellow shaded regions indicate the 3 and 5$\sigma$ uncertainties in photometric error. The black dashed line denotes the level of the main sequence turnoff of the oldest population observed in Liller 1. The open circle corresponds to the position of the blend between the RBC and the nearby RGB star, while the red circle corresponds to the resolved RGB star, firmly confirming that it is on the red giant branch. Stars have been proper-motion selected and the magnitudes have been corrected for differential reddening.}
    \label{fig:cmd}
\end{figure}

We refer to the color-magnitude diagram (CMD) produced and described in detail in \cite{2021NatAs...5..311F} and \cite{2022ApJ...940..170D}. The CMD shown in Figure~\ref{fig:cmd} contains proper-motion verified cluster members and has been differentially corrected for reddening with the methodology in \cite{2021ApJ...917...92P}. Overplotted on the CMD are 1, 2, and 12 Gyr isochrones which map stellar evolution tracks representative of the young and old populations contained in Liller 1. Using \texttt{DAOPhot}'s photometric capabilities, we have deblended the two sources and confirm that the northern source is located on the red giant branch and we provide a F184W magnitude constraint on the RB counterpart. If the donor is part of the old population, then this constraint suggests that it is a possible main-sequence turnoff star.

\section{Results} \label{sec:results}

\subsection{Localization}

We report the updated coordinate of the RBC at an RA, Dec of (17h33m24.5874s, $-$33d23m20.259s), (ICRS, epoch 2025.31; Table~\ref{tab:rb_positions}). As Figure~\ref{fig:localization} shows, our localization is consistent with the 1-$\sigma$ error ellipse from the updated Chandra Source Catalog 2.1 (CSC2.1; \citealp{2024ApJS..274...22E}), as well as the proper-motion corrected original X-ray localization. We note a modest offset between the radio localization and the position of our source, with our localization falling within $3\sigma$ of the radio error ellipse. We also identify P25's source in the JWST observations. Though it falls within the $3\sigma$ X-ray localization from \cite{2001AJ....122.2627H}, it is inconsistent with the $1\sigma$ errors from the X-ray and radio constraints.

\subsection{Pallanca et al. 2025's Source}

We compute a Lomb--Scargle periodogram for the P25 candidate and find the best period with no sinusoidal residuals is at 1068 mins (17.8 hours). With a 34.82 hour gap between subsequent observations, this allows for nearly 2 cycles to pass. The resulting phase-folded light curve allows us to classify this source as a likely contact binary where both stars have filled their Roche lobes and share an outer envelope. Contact binary light curves are smooth and show two distinct eclipses of similar depth, with no time spent out of eclipse. The close separation between the two stars results in a short sub-day period, consistent with our measurement.

\subsection{CMD Results}

Our photometry provides us with reliable measurements in the F200W wavelength band, but the unresolved blend on the NRCBLONG detector prevents us from reporting an uncontaminated F356W measurement. Turning to previous observations, we isolate the photometry of the non-variable component of the blend in the Ks and F814W bands. Placing it on the CMD shown in Figure~\ref{fig:cmd}, we confirm that this star is on the red giant branch. Since the RB was in quiescence during the Gemini and HST observations, we only faithfully recover a photometric measurement in the F814W band. This allows us to place a magnitude limit on the counterpart, but the lack of color information prevents us from accurately reporting its physical properties including its effective temperature, age, and surface gravity. 


\subsection{Discussion and Constraints}

It is unlikely that the red giant is gravitationally bound to the system. At a distance of 8.5 kpc, the lower limit on their separation is $\sim$ 360 AU. If we assume a binary of 1.4 M$_\odot$ and 1 M$_\odot$, their orbital velocity at 360 AU is 2.43 km/s. The limit between a hard and a soft binary occurs when the binding energy exceeds the typical kinetic energy of a passing cluster star:

\begin{equation}
    a_{\text{hs}} = \frac{Gm_1m_2}{\langle m \rangle \sigma_{RV}^2}.
\end{equation}

\noindent Taking an average cluster star to have a mass $\langle m \rangle = 0.5 \text{M}_\odot$ and Liller 1's velocity dispersion to be $\sigma_{RV} = \sim 20$ km/s \citep{2019MNRAS.482.5138B}, we arrive at a separation $a_{\text{hs}} = 6.2$ AU for the pair to remain bound. This is far below the lower limit of 360 AU, so any chance encounters with other stars in Liller 1's core would drive the two stars apart. This indicates that the red giant is not gravitationally bound to the RB, nor can it be the donor to the system.  

\begin{figure}
    \centering
    \includegraphics[width=\linewidth]{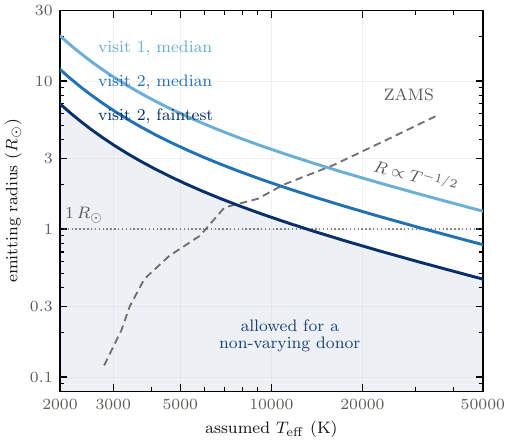}
    \caption{Upper limit on the radius of the RB donor as a function of its assumed effective temperature, for a spherical blackbody at $d = 8.5$ kpc. Curves give the radius required to reproduce the observed dereddened F200W flux in the visit 1 median (light), visit 2 median (middle) and faintest visit 2 (dark) states. Because the donor cannot vary while the disk can, the faintest state sets the limit and the shaded region below it is allowed. The dashed line is an approximate ZAMS locus, shown for orientation only. The limit scales as $R \propto d$ and as $R \propto 10^{-0.2 m_{0}}$.}
    \label{fig:radii}
\end{figure}

We can place an upper limit on the radius of the donor in the RB system. The donor's photosphere cannot vary over the span of these observations, whereas the accretion disk can, so the faintest state recorded in visit 2 ($m_{0} = 16.73$) bounds the donor's contribution from above. Attributing that entire flux to the donor and treating it as a sphere radiating as a blackbody at temperature $T_{\rm eff}$ gives

\begin{equation}
R = d \left[ \frac{F_\nu}{\pi B_\nu(T_{\rm eff})} \right]^{1/2},
\end{equation}

\noindent shown as the darkest curve in Figure~\ref{fig:radii}. Adopting $d = 8.5$ kpc, the limit ranges from $R < 3.8\,R_\odot$ at 3000 K to $R < 0.78\,R_\odot$ at 20,000 K, following $R \propto T_{\rm eff}^{-1/2}$ above $\sim$7000 K where the band lies in the Rayleigh--Jeans regime. A zero-age main-sequence star would exceed this limit for $T_{\rm eff} \gtrsim 7700$ K, excluding an unevolved donor earlier than roughly an A spectral type. Any disk contribution to the faint state reduces the limit by $(1-f_{\rm disk})^{1/2}$, and the limit scales linearly with the assumed distance.

\section{Conclusion}\label{sec:conc}

We present the first glimpse at the counterpart to the RB, providing new coordinates and photometry of the source. The source is blended with a red giant star with an angular separation of around 42 mas; still, we are able to recover the F200W light curve of the RB counterpart. 

Simultaneous NIR-X-ray observations are important in isolating X-ray and IR bursts from the RB. The Monitor of All-sky X-ray Image (MAXI, \citep{2009PASJ...61..999M}) telescope actively monitored the region containing the Rapid Burster during our JWST observations. The MAXI light curve centered on the Rapid Burster coordinates (see Appendix, Figure~\ref{fig:maxi}) confirms that the Rapid Burster was in outburst. This outburst reaches a 2--10~keV luminosity of $L_X\approx8 \times 10^{36}\,\text{erg s}^{-1}$, approaching the 2023--2025 average outburst peak of $\sim1 \times 10^{37}\,\text{erg s}^{-1}$ \citep{2025ApJS..279...57H}. While the flares present in the second visit of the light curve of the RB counterpart shown in Figure~\ref{fig:lcs} demonstrate coherent and astrophysical variability of the source, we cannot align these changes in brightness with the flaring behavior of the RB without higher time resolution concurrent X-ray observations. These observations would illuminate if there is potential reprocessing of X-ray emission in the accretion disk or the donor star; the latter was the case for Type I flares in the 4U 1728-34 neutron star system \cite{2020MNRAS.495L..37V}. This necessitates future observation of time lags across different wavelengths. Spectroscopy will also be a vital tool in understanding the morphology of the system's accretion disk and potentially classifying the system's donor.

While we successfully recover the F814W and F200W light curves, these observations were taken at different times and during different outburst states. However, this indicates that concurrent optical and infrared observations could also provide a photometric constraint on the age of the counterpart. Instruments used to perform this task should be selected for the highest pixel resolution available to combat crowding issues to the furthest extent. 

The implications of this work indicate that Liller 1's dense core has the ability to produce rare sources. As the RB accretes mass through Roche-Lobe overflow, angular momentum is transferred to the neutron star and its rotation rate increases \citep{1982Natur.300..728A}. If accretion onto the star ceases, it is expected that the RB will evolve to become a millisecond pulsar. Currently, no millisecond pulsars have been confirmed in Liller 1 even though it provides an ideal environment for their development \citep{2024ApJ...969L...7Y}. Additionally, Liller 1 is known to be gamma-ray bright, which can be explained by a large population of unresolved pulsars within the cluster \citep{2011ApJ...729...90T}. 

The lack of observed pulsars may be due to the comparatively limited X-ray observations of Liller 1, made difficult by the high neutral hydrogen column density $N_H$, which limits the detectability of X-ray sources to those with fluxes above 1 keV \citep{2017arXiv170307221I, 2012ApJ...752..158S}. When compared to Terzan 5, which exhibits a similarly complex stellar population \citep{ferraro+09, ferraro+16, massari+14, origlia+25, zullo+26}, Liller 1 lags in detected sources. In fact,  Terzan 5 is known to host 15\% of all the  millisecond pulsars observed in Milky Way globular clusters \citep{ransom+05, cadelano+18, padmanabh+24}. These dense stellar systems are expected to form an excess of compact object binaries through dynamical interactions, and the lack of known systems in Liller 1 is surprising \citep{2007A&G....48e..12M, 2026enap....3..458K}. In this new era of high-precision imaging, this motivates further study of Liller 1's core.

\begin{acknowledgments}
M.M.D. acknowledges Herman L. Marshall, Joheen Chakraborty, and Vera L. Berger for their valuable time and expertise spent discussing various topics related to the RB, X-ray data, and optical photometry. This work is based on observations made with the NASA/ESA/CSA James Webb Space Telescope. The data were obtained from the Mikulski Archive for Space Telescopes at the Space Telescope Science Institute, which is operated by the Association of Universities for Research in Astronomy, Inc., under NASA contract NAS5-03127 for JWST. These observations are associated with program \#5381. M.D. and K.B.B. acknowledge support from NASA through grant JWST-GO-05381.005-A from the Space Telescope Science Institute, which is operated by the Association of Universities for Research in Astronomy, Inc., under NASA contract NAS5-03127. F.R.F., B.L. and C.P. acknowledge financial support from the project GENESIS – Searching for the primordial structures of the Universe in the heart of the Galaxy (Advanced Grant FIS-2024-02056, PI:Ferraro), funded by the Italian MUR through the Fondo Italiano per la Scienza  (FIS3) call. M.N. is a Fonds de Recherche du Quebec – Nature et Technologies (FRQNT) postdoctoral fellow. This research is based in part on observations made with the NASA/ESA Hubble Space Telescope obtained from the Space Telescope Science Institute, which is operated by the Association of Universities for Research in Astronomy, Inc., under NASA contract NAS 5–26555. These observations are associated with program 15231. This research has made use of the MAXI data provided by RIKEN, JAXA and the MAXI team.

\end{acknowledgments}

\appendix

For posterity, we provide median-subtracted snapshots of the 192 zeroframe images used across both segments in both JWST wavelength bands centered on the RBC. Changes in brightness appear in both wavelength bands at the same time, and several panels indicate rapid variability from frame to frame.  

\begin{figure*}
    \centering
    \includegraphics[width=0.85\linewidth]{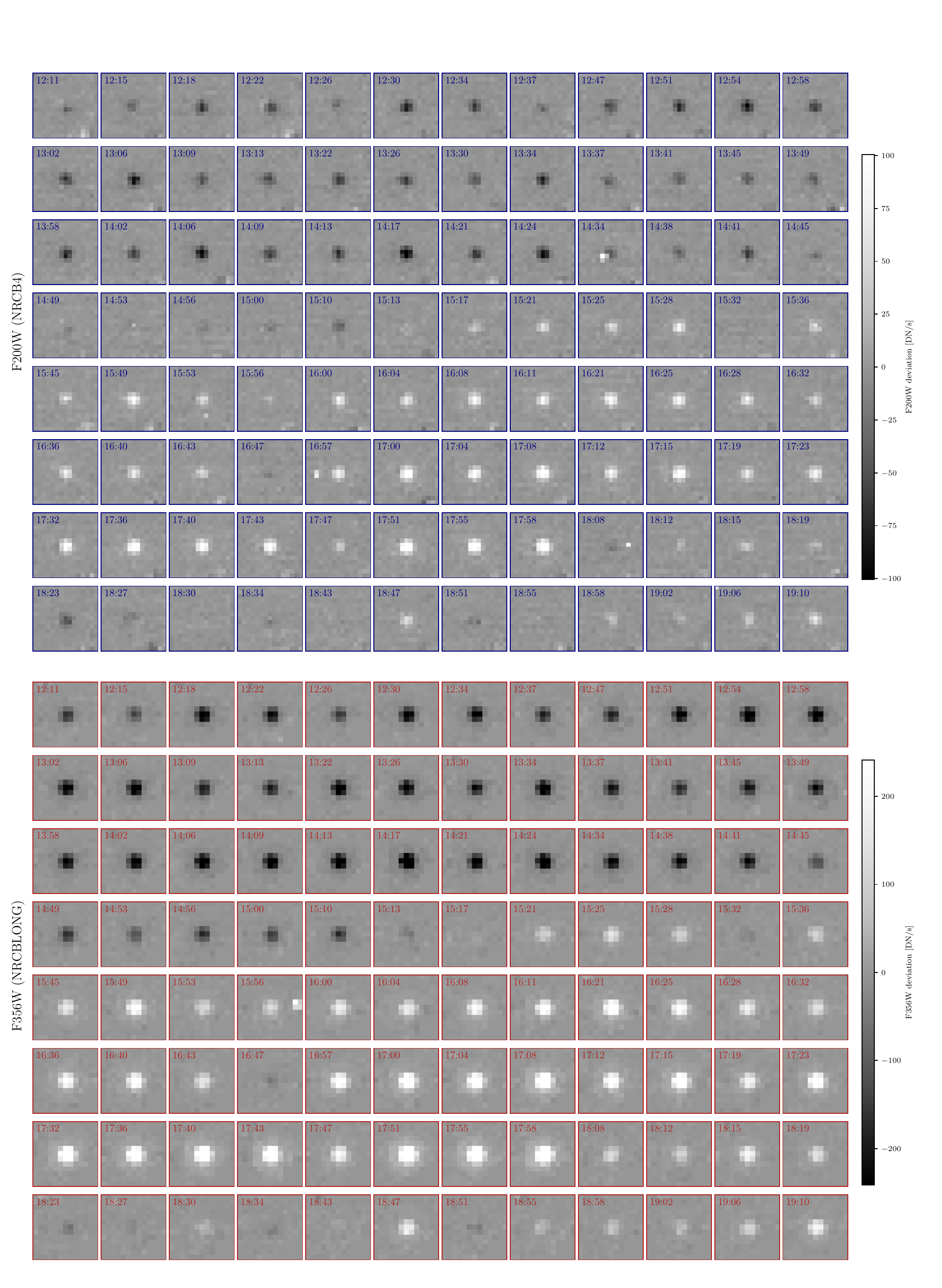}
    \caption{Rapid Burster counterpart evolution over the first data segment in the F200W (top, blue outline) and F356W (bottom, red outline) bands. Time is shown in the top left corner, running from April 22, 2025 12:11 to 19:10 UTC.}
    \label{fig:collageseg3}
\end{figure*}

\begin{figure*}
    \centering
    \includegraphics[width=0.85\linewidth]{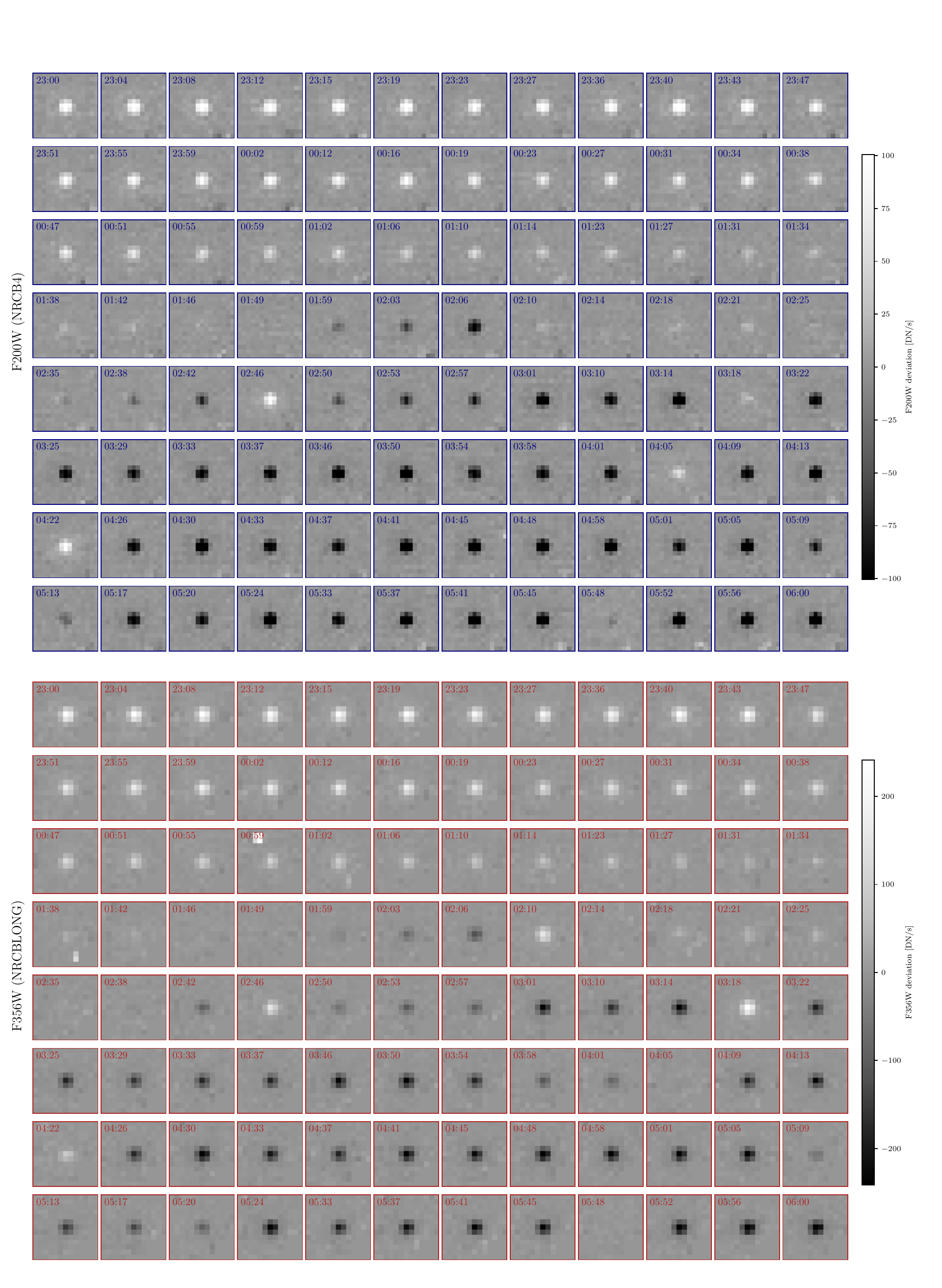}
    \caption{Rapid Burster counterpart evolution over the second data segment in the F200W (top, blue outline) and F356W (bottom, red outline) bands. Time is shown in the top left corner, running from April 23, 2025 23:00 to April 24, 2025 06:00 UTC.}
    \label{fig:collageseg4}
\end{figure*}

\begin{figure*}
    \centering
    \includegraphics[width=0.85\linewidth]{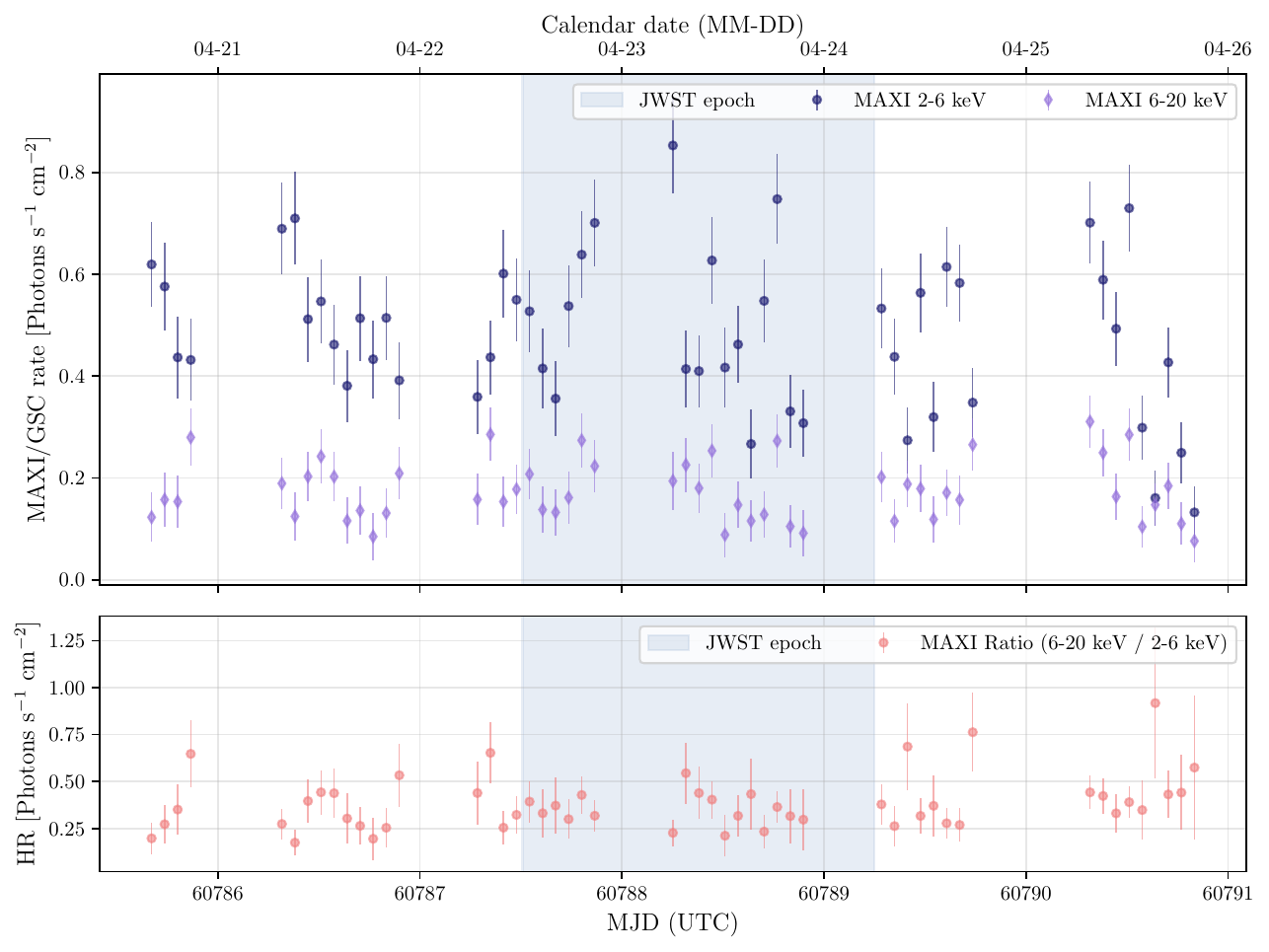}
    \caption{MAXI data centered on the Rapid Burster. Top panel: MAXI light curve, binned in one orbit increments, with soft X-rays (2--6~keV) displayed as dark purple circles and hard X-rays (6--20~keV) displayed as light purple diamonds. Bottom panel: MAXI hardness ratio.}
    \label{fig:maxi}
\end{figure*}

We display here the MAXI data available during the week the JWST observations were taken in Figure~\ref{fig:maxi}. The X-ray data, while contaminated with data from the X-ray active neutron star GX 354$-$0 (the Slow Burster) 1.5 degrees away, shows a high level of X-ray activity from the region centered around the Rapid Burster.

\bibliography{biblio}

@ARTICLE{padmanabh+24,
       author = {{Padmanabh}, P.~V. and {Ransom}, S.~M. and {Freire}, P.~C.~C. and {Ridolfi}, A. and {Taylor}, J.~D. and {Choza}, C. and {Clark}, C.~J. and {Abbate}, F. and {Bailes}, M. and {Barr}, E.~D. and {Buchner}, S. and {Burgay}, M. and {DeCesar}, M.~E. and {Chen}, W. and {Corongiu}, A. and {Champion}, D.~J. and {Dutta}, A. and {Geyer}, M. and {Hessels}, J.~W.~T. and {Kramer}, M. and {Possenti}, A. and {Stairs}, I.~H. and {Stappers}, B.~W. and {Venkatraman Krishnan}, V. and {Vleeschower}, L. and {Zhang}, L.},
        title = "{Discovery and timing of ten new millisecond pulsars in the globular cluster Terzan 5}",
      journal = {\aap},
         year = 2024,
        month = jun,
       volume = {686},
          eid = {A166},
        pages = {A166},
          doi = {10.1051/0004-6361/202449303},
archivePrefix = {arXiv},
       eprint = {2403.17799},
 primaryClass = {astro-ph.HE},
       adsurl = {https://ui.adsabs.harvard.edu/abs/2024A&A...686A.166P}
}

@ARTICLE{cadelano+18,
       author = {{Cadelano}, M. and {Ransom}, S.~M. and {Freire}, P.~C.~C. and {Ferraro}, F.~R. and {Hessels}, J.~W.~T. and {Lanzoni}, B. and {Pallanca}, C. and {Stairs}, I.~H.},
        title = "{Discovery of Three New Millisecond Pulsars in Terzan 5}",
      journal = {\apj},
         year = 2018,
        month = mar,
       volume = {855},
       number = {2},
          eid = {125},
        pages = {125},
          doi = {10.3847/1538-4357/aaac2a},
archivePrefix = {arXiv},
       eprint = {1801.09929},
 primaryClass = {astro-ph.HE},
       adsurl = {https://ui.adsabs.harvard.edu/abs/2018ApJ...855..125C}
}

@ARTICLE{ransom+05,
       author = {{Ransom}, Scott M. and {Hessels}, Jason W.~T. and {Stairs}, Ingrid H. and {Freire}, Paulo C.~C. and {Camilo}, Fernando and {Kaspi}, Victoria M. and {Kaplan}, David L.},
        title = "{Twenty-One Millisecond Pulsars in Terzan 5 Using the Green Bank Telescope}",
      journal = {Science},
         year = 2005,
        month = feb,
       volume = {307},
       number = {5711},
        pages = {892-896},
          doi = {10.1126/science.1108632},
archivePrefix = {arXiv},
       eprint = {astro-ph/0501230},
 primaryClass = {astro-ph},
       adsurl = {https://ui.adsabs.harvard.edu/abs/2005Sci...307..892R}
}

@ARTICLE{chiappino+26,
       author = {{Chiappino}, L. and {Origlia}, L. and {Fanelli}, C. and {Bartolomei}, A. and {Ferraro}, F.~R. and {Lanzoni}, B. and {Pallanca}, C. and {Cadelano}, M. and {Romano}, D. and {Dalessandro}, E. and {Massari}, D. and {Valenti}, E. and {Rich}, R.~M.},
        title = "{CRIRES+ reveals the chemistry of the stellar sub-populations in the bulge fossil fragment Liller 1}",
      journal = {arXiv e-prints},
         year = 2026,
        month = jun,
          eid = {arXiv:2606.11329},
        pages = {arXiv:2606.11329},
          doi = {10.48550/arXiv.2606.11329},
archivePrefix = {arXiv},
       eprint = {2606.11329},
 primaryClass = {astro-ph.GA},
       adsurl = {https://ui.adsabs.harvard.edu/abs/2026arXiv260611329C}
}

@ARTICLE{fanelli+24,
       author = {{Fanelli}, C. and {Origlia}, L. and {Rich}, R.~M. and {Ferraro}, F.~R. and {Alvarez Garay}, D.~A. and {Chiappino}, L. and {Lanzoni}, B. and {Pallanca}, C. and {Crociati}, C. and {Dalessandro}, E.},
        title = "{Multi-iron subpopulations in Liller 1 from high-resolution H-band spectroscopy}",
      journal = {\aap},
         year = 2024,
        month = oct,
       volume = {690},
          eid = {A139},
        pages = {A139},
          doi = {10.1051/0004-6361/202451030},
archivePrefix = {arXiv},
       eprint = {2408.12649},
 primaryClass = {astro-ph.GA},
       adsurl = {https://ui.adsabs.harvard.edu/abs/2024A&A...690A.139F}
}

@ARTICLE{deimer+24,
       author = {{Alvarez Garay}, D.~A. and {Fanelli}, C. and {Origlia}, L. and {Pallanca}, C. and {Mucciarelli}, A. and {Chiappino}, L. and {Crociati}, C. and {Lanzoni}, B. and {Ferraro}, F.~R. and {Rich}, R.~M. and {Dalessandro}, E.},
        title = "{X-shooter spectroscopy of Liller 1 giant stars}",
      journal = {\aap},
         year = 2024,
        month = jun,
       volume = {686},
          eid = {A198},
        pages = {A198},
          doi = {10.1051/0004-6361/202449595},
archivePrefix = {arXiv},
       eprint = {2404.14130},
 primaryClass = {astro-ph.GA},
       adsurl = {https://ui.adsabs.harvard.edu/abs/2024A&A...686A.198A}
}

@ARTICLE{crociati+23,
       author = {{Crociati}, Chiara and {Valenti}, Elena and {Ferraro}, Francesco R. and {Pallanca}, Cristina and {Lanzoni}, Barbara and {Cadelano}, Mario and {Fanelli}, Cristiano and {Origlia}, Livia and {Leanza}, Silvia and {Dalessandro}, Emanuele and {Mucciarelli}, Alessio and {Rich}, R. Michael},
        title = "{First Evidence of Multi-iron Subpopulations in the Bulge Fossil Fragment Candidate Liller 1}",
      journal = {\apj},
         year = 2023,
        month = jul,
       volume = {951},
       number = {1},
          eid = {17},
        pages = {17},
          doi = {10.3847/1538-4357/acd382},
archivePrefix = {arXiv},
       eprint = {2305.04595},
 primaryClass = {astro-ph.GA},
       adsurl = {https://ui.adsabs.harvard.edu/abs/2023ApJ...951...17C}
}

@ARTICLE{zullo+26,
       author = {{Zullo}, G. and {Pallanca}, C. and {Ferraro}, F.~R. and {Lanzoni}, B. and {Origlia}, L. and {Massari}, D. and {Dalessandro}, E. and {Fanelli}, C. and {Cadelano}, M. and {Vesperini}, E. and {Crociati}, C. and {Rich}, R.~M. and {Valenti}, E.},
        title = "{The multi-age stellar populations of Terzan 5 as revealed by JWST}",
      journal = {\aap},
         year = 2026,
        month = may,
       volume = {709},
          eid = {A212},
        pages = {A212},
          doi = {10.1051/0004-6361/202659349},
archivePrefix = {arXiv},
       eprint = {2604.00098},
 primaryClass = {astro-ph.GA},
       adsurl = {https://ui.adsabs.harvard.edu/abs/2026A&A...709A.212Z}
}

@ARTICLE{origlia+25,
       author = {{Origlia}, L. and {Ferraro}, F.~R. and {Fanelli}, C. and {Lanzoni}, B. and {Massari}, D. and {Dalessandro}, E. and {Pallanca}, C.},
        title = "{The manifest link between Terzan 5 and the Galactic bulge}",
      journal = {\aap},
         year = 2025,
        month = may,
       volume = {697},
          eid = {A19},
        pages = {A19},
          doi = {10.1051/0004-6361/202452110},
archivePrefix = {arXiv},
       eprint = {2503.17258},
 primaryClass = {astro-ph.GA},
       adsurl = {https://ui.adsabs.harvard.edu/abs/2025A&A...697A..19O}
}

@ARTICLE{ferraro+25,
       author = {{Ferraro}, F.~R. and {Chiappino}, L. and {Bartolomei}, A. and {Origlia}, L. and {Fanelli}, C. and {Lanzoni}, B. and {Pallanca}, C. and {Loriga}, M. and {Leanza}, S. and {Valenti}, E. and {Romano}, D. and {Mucciarelli}, A. and {Massari}, D. and {Cadelano}, M. and {Dalessandro}, E. and {Crociati}, C. and {Rich}, R.~M.},
        title = "{The Bulge Cluster Origin (BulCO) survey at the ESO-VLT: Probing the early history of the Milky Way assembly. Design and first results in Liller 1}",
      journal = {\aap},
         year = 2025,
        month = apr,
       volume = {696},
          eid = {A179},
        pages = {A179},
          doi = {10.1051/0004-6361/202554092},
archivePrefix = {arXiv},
       eprint = {2503.14642},
 primaryClass = {astro-ph.GA},
       adsurl = {https://ui.adsabs.harvard.edu/abs/2025A&A...696A.179F}
}

@ARTICLE{dalessandro+22,
       author = {{Dalessandro}, Emanuele and {Crociati}, Chiara and {Cignoni}, Michele and {Ferraro}, Francesco R. and {Lanzoni}, Barbara and {Origlia}, Livia and {Pallanca}, Cristina and {Rich}, R. Michael and {Saracino}, Sara and {Valenti}, Elena},
        title = "{Clues to the Formation of Liller 1 from Modeling Its Complex Star Formation History}",
      journal = {\apj},
         year = 2022,
        month = dec,
       volume = {940},
       number = {2},
          eid = {170},
        pages = {170},
          doi = {10.3847/1538-4357/ac9907},
archivePrefix = {arXiv},
       eprint = {2210.05694},
 primaryClass = {astro-ph.GA},
       adsurl = {https://ui.adsabs.harvard.edu/abs/2022ApJ...940..170D}
}

@ARTICLE{ferraro+21,
       author = {{Ferraro}, F.~R. and {Pallanca}, C. and {Lanzoni}, B. and {Crociati}, C. and {Dalessandro}, E. and {Origlia}, L. and {Rich}, R.~M. and {Saracino}, S. and {Mucciarelli}, A. and {Valenti}, E. and {Geisler}, D. and {Mauro}, F. and {Villanova}, S. and {Moni Bidin}, C. and {Beccari}, G.},
        title = "{A new class of fossil fragments from the hierarchical assembly of the Galactic bulge}",
      journal = {Nature Astronomy},
         year = 2021,
        month = jan,
       volume = {5},
        pages = {311-318},
          doi = {10.1038/s41550-020-01267-y},
archivePrefix = {arXiv},
       eprint = {2011.09966},
 primaryClass = {astro-ph.GA},
       adsurl = {https://ui.adsabs.harvard.edu/abs/2021NatAs...5..311F}
}

@ARTICLE{ferraro+16,
       author = {{Ferraro}, F.~R. and {Massari}, D. and {Dalessandro}, E. and {Lanzoni}, B. and {Origlia}, L. and {Rich}, R.~M. and {Mucciarelli}, A.},
        title = "{The Age of the Young Bulge-like Population in the Stellar System Terzan 5: Linking the Galactic Bulge to the High-z Universe}",
      journal = {\apj},
         year = 2016,
        month = sep,
       volume = {828},
       number = {2},
          eid = {75},
        pages = {75},
          doi = {10.3847/0004-637X/828/2/75},
archivePrefix = {arXiv},
       eprint = {1609.01515},
 primaryClass = {astro-ph.GA},
       adsurl = {https://ui.adsabs.harvard.edu/abs/2016ApJ...828...75F}
}

@ARTICLE{massari+14,
       author = {{Massari}, D. and {Mucciarelli}, A. and {Ferraro}, F.~R. and {Origlia}, L. and {Rich}, R.~M. and {Lanzoni}, B. and {Dalessandro}, E. and {Valenti}, E. and {Ibata}, R. and {Lovisi}, L. and {Bellazzini}, M. and {Reitzel}, D.},
        title = "{Ceci N'est Pas a Globular Cluster: The Metallicity Distribution of the Stellar System Terzan 5}",
      journal = {\apj},
         year = 2014,
        month = nov,
       volume = {795},
       number = {1},
          eid = {22},
        pages = {22},
          doi = {10.1088/0004-637X/795/1/22},
archivePrefix = {arXiv},
       eprint = {1409.1682},
 primaryClass = {astro-ph.SR},
       adsurl = {https://ui.adsabs.harvard.edu/abs/2014ApJ...795...22M}
}

@ARTICLE{ferraro+09,
       author = {{Ferraro}, F.~R. and {Dalessandro}, E. and {Mucciarelli}, A. and {Beccari}, G. and {Rich}, R.~M. and {Origlia}, L. and {Lanzoni}, B. and {Rood}, R.~T. and {Valenti}, E. and {Bellazzini}, M. and {Ransom}, S.~M. and {Cocozza}, G.},
        title = "{The cluster Terzan 5 as a remnant of a primordial building block of the Galactic bulge}",
      journal = {\nat},
         year = 2009,
        month = nov,
       volume = {462},
       number = {7272},
        pages = {483-486},
          doi = {10.1038/nature08581},
archivePrefix = {arXiv},
       eprint = {0912.0192},
 primaryClass = {astro-ph.GA},
       adsurl = {https://ui.adsabs.harvard.edu/abs/2009Natur.462..483F}
}

@ARTICLE{1978Natur.271..630H,
       author = {{Hoffman}, J.~A. and {Marshall}, H.~L. and {Lewin}, W.~H.~G.},
        title = "{Dual character of the rapid burster and a classification of X-ray bursts}",
      journal = {\nat},
         year = 1978,
        month = feb,
       volume = {271},
       number = {5646},
        pages = {630-633},
          doi = {10.1038/271630a0},
       adsurl = {https://ui.adsabs.harvard.edu/abs/1978Natur.271..630H}
}

@ARTICLE{1976ApJ...210L..13H,
       author = {{Hoffman}, J.~A. and {Lewin}, W.~H.~G. and {Doty}, J. and {Hearn}, D.~R. and {Clark}, G.~W. and {Jernigan}, G. and {Li}, F.~K.},
        title = "{Discovery of X-ray bursts from MXB 1728-34.}",
      journal = {\apjl},
         year = 1976,
        month = nov,
       volume = {210},
        pages = {L13-L17},
          doi = {10.1086/182292},
       adsurl = {https://ui.adsabs.harvard.edu/abs/1976ApJ...210L..13H}
}

@ARTICLE{1976ApJ...207L..95L,
       author = {{Lewin}, W.~H.~G. and {Doty}, J. and {Clark}, G.~W. and {Rappaport}, S.~A. and {Bradt}, H.~V.~D. and {Doxsey}, R. and {Hearn}, D.~R. and {Hoffman}, J.~A. and {Jernigan}, J.~G. and {Li}, F.~K. and {Mayer}, W. and {McClintock}, J. and {Primini}, F. and {Richardson}, J.},
        title = "{The discovery of rapidly repetitive X-ray bursts from a new source in Scorpius.}",
      journal = {\apjl},
         year = 1976,
        month = jul,
       volume = {207},
        pages = {L95-L99},
          doi = {10.1086/182188},
       adsurl = {https://ui.adsabs.harvard.edu/abs/1976ApJ...207L..95L}
}

@INPROCEEDINGS{2021ASSL..461..209G,
       author = {{Galloway}, Duncan K. and {Keek}, Laurens},
        title = "{Thermonuclear X-ray Bursts}",
    booktitle = {Timing Neutron Stars: Pulsations, Oscillations and Explosions},
         year = 2021,
       editor = {{Belloni}, Tomaso M. and {M{\'e}ndez}, Mariano and {Zhang}, Chengmin},
       series = {Astrophysics and Space Science Library},
       volume = {461},
        month = jan,
        pages = {209-262},
          doi = {10.1007/978-3-662-62110-3_5},
archivePrefix = {arXiv},
       eprint = {1712.06227},
 primaryClass = {astro-ph.HE},
       adsurl = {https://ui.adsabs.harvard.edu/abs/2021ASSL..461..209G}
}

@ARTICLE{2020ApJS..249...32G,
       author = {{Galloway}, Duncan K. and {in't Zand}, Jean and {Chenevez}, J{\'e}r{\^o}me and {W{\"o}rpel}, Hauke and {Keek}, Laurens and {Ootes}, Laura and {Watts}, Anna L. and {Gisler}, Luis and {Sanchez-Fernandez}, Celia and {Kuulkers}, Erik},
        title = "{The Multi-INstrument Burst ARchive (MINBAR)}",
      journal = {\apjs},
         year = 2020,
        month = aug,
       volume = {249},
       number = {2},
          eid = {32},
        pages = {32},
          doi = {10.3847/1538-4365/ab9f2e},
archivePrefix = {arXiv},
       eprint = {2003.00685},
 primaryClass = {astro-ph.HE},
       adsurl = {https://ui.adsabs.harvard.edu/abs/2020ApJS..249...32G}
}

@ARTICLE{1993SSRv...62..223L,
       author = {{Lewin}, Walter H.~G. and {van Paradijs}, Jan and {Taam}, Ronald E.},
        title = "{X-Ray Bursts}",
      journal = {\ssr},
         year = 1993,
        month = sep,
       volume = {62},
       number = {3-4},
        pages = {223-389},
          doi = {10.1007/BF00196124},
       adsurl = {https://ui.adsabs.harvard.edu/abs/1993SSRv...62..223L}
}

@ARTICLE{1996Natur.381..291F,
       author = {{Finger}, Mark H. and {Koh}, Danny T. and {Nelson}, Robert W. and {Prince}, Thomas A. and {Vaughan}, Brian A. and {Wilson}, Robert B.},
        title = "{Discovery of hard X-ray pulsations from the transient source GRO J1744 - 28}",
      journal = {\nat},
         year = 1996,
        month = may,
       volume = {381},
       number = {6580},
        pages = {291-293},
          doi = {10.1038/381291a0},
       adsurl = {https://ui.adsabs.harvard.edu/abs/1996Natur.381..291F}
}

@ARTICLE{1999MNRAS.307..179G,
       author = {{Guerriero}, R. and {Fox}, D.~W. and {Kommers}, J. and {Lewin}, W.~H.~G. and {Rutledge}, R. and {Moore}, C.~B. and {Morgan}, E. and {van Paradijs}, J. and {van der Klis}, M. and {Bildsten}, L. and {Dotani}, T.},
        title = "{The evolution of Rapid Burster outbursts}",
      journal = {\mnras},
         year = 1999,
        month = jul,
       volume = {307},
       number = {1},
        pages = {179-189},
          doi = {10.1046/j.1365-8711.1999.02651.x},
archivePrefix = {arXiv},
       eprint = {astro-ph/9807110},
 primaryClass = {astro-ph},
       adsurl = {https://ui.adsabs.harvard.edu/abs/1999MNRAS.307..179G}
}

@ARTICLE{2001AJ....122.2627H,
       author = {{Homer}, L. and {Deutsch}, Eric W. and {Anderson}, Scott F. and {Margon}, Bruce},
        title = "{The Rapid Burster in Liller 1: The Chandra X-Ray Position and a Search for an Infrared Counterpart}",
      journal = {\aj},
         year = 2001,
        month = nov,
       volume = {122},
       number = {5},
        pages = {2627-2633},
          doi = {10.1086/323545},
archivePrefix = {arXiv},
       eprint = {astro-ph/0106140},
 primaryClass = {astro-ph},
       adsurl = {https://ui.adsabs.harvard.edu/abs/2001AJ....122.2627H}
}

@ARTICLE{2021ApJ...917...92P,
       author = {{Pallanca}, Cristina and {Ferraro}, Francesco R. and {Lanzoni}, Barbara and {Crociati}, Chiara and {Saracino}, Sara and {Dalessandro}, Emanuele and {Origlia}, Livia and {Rich}, Michael R. and {Valenti}, Elena and {Geisler}, Douglas and {Mauro}, Francesco and {Villanova}, Sandro and {Moni Bidin}, Christian and {Beccari}, Giacomo},
        title = "{High-resolution Extinction Map in the Direction of the Strongly Obscured Bulge Fossil Fragment Liller 1}",
      journal = {\apj},
         year = 2021,
        month = aug,
       volume = {917},
       number = {2},
          eid = {92},
        pages = {92},
          doi = {10.3847/1538-4357/ac0889},
archivePrefix = {arXiv},
       eprint = {2106.02448},
 primaryClass = {astro-ph.GA},
       adsurl = {https://ui.adsabs.harvard.edu/abs/2021ApJ...917...92P}
}

@article{2018RenormalisingTA,
  title={Re-normalising the astrometric chi-square in Gaia DR2},
  author={Lindegren, L.},
  howpublished={Gaia Technical Note},
  year={2018},
  url={https://api.semanticscholar.org/CorpusID:195836829}
}

@ARTICLE{2015MNRAS.449..268B,
       author = {{Bagnoli}, T. and {in't Zand}, J.~J.~M. and {D'Angelo}, C.~R. and {Galloway}, D.~K.},
        title = "{A population study of type II bursts in the Rapid Burster}",
      journal = {\mnras},
         year = 2015,
        month = may,
       volume = {449},
       number = {1},
        pages = {268-287},
          doi = {10.1093/mnras/stv330},
archivePrefix = {arXiv},
       eprint = {1502.03941},
 primaryClass = {astro-ph.HE},
       adsurl = {https://ui.adsabs.harvard.edu/abs/2015MNRAS.449..268B}
}

@ARTICLE{Pallanca2025,
       author = {{Pallanca}, Cristina and {Ferraro}, Francesco R. and {Lanzoni}, Barbara and {Cadelano}, Mario and {Heinke}, Craig O. and {van den Berg}, Maureen and {Homan}, Jeroen and {Crociati}, Chiara and {Guillot}, Sebastien},
        title = "{Potential discovery of the long-sought optical counterpart to the Rapid Burster in the bulge fossil fragment Liller 1}",
      journal = {\aap},
         year = 2025,
        month = nov,
       volume = {703},
          eid = {A182},
        pages = {A182},
          doi = {10.1051/0004-6361/202556238},
archivePrefix = {arXiv},
       eprint = {2509.10640},
 primaryClass = {astro-ph.SR},
       adsurl = {https://ui.adsabs.harvard.edu/abs/2025A&A...703A.182P}
}

@misc{ferraro2019_hst15231,
  author       = {Ferraro, Francesco R.},
  title        = {{HST} Proposal 15231},
  year         = {2019},
  publisher    = {Mikulski Archive for Space Telescopes},
  doi          = {10.17909/7zah-te68},
  url          = {https://doi.org/10.17909/7zah-te68}
}

@ARTICLE{2021NatAs...5..311F,
       author = {{Ferraro}, F.~R. and {Pallanca}, C. and {Lanzoni}, B. and {Crociati}, C. and {Dalessandro}, E. and {Origlia}, L. and {Rich}, R.~M. and {Saracino}, S. and {Mucciarelli}, A. and {Valenti}, E. and {Geisler}, D. and {Mauro}, F. and {Villanova}, S. and {Moni Bidin}, C. and {Beccari}, G.},
        title = "{A new class of fossil fragments from the hierarchical assembly of the Galactic bulge}",
      journal = {Nature Astronomy},
         year = 2021,
        month = jan,
       volume = {5},
        pages = {311-318},
          doi = {10.1038/s41550-020-01267-y},
archivePrefix = {arXiv},
       eprint = {2011.09966},
 primaryClass = {astro-ph.GA},
       adsurl = {https://ui.adsabs.harvard.edu/abs/2021NatAs...5..311F}
}

@ARTICLE{1995AJ....109.1154F,
       author = {{Frogel}, Jay A. and {Kuchinski}, Leslie E. and {Tiede}, Glenn P.},
        title = "{Infrared Array Photometry of Metal-Rich Globular Clusters. II. Liller 1- The Most Metal Rich Cluster?}",
      journal = {\aj},
         year = 1995,
        month = mar,
       volume = {109},
        pages = {1154},
          doi = {10.1086/117348},
       adsurl = {https://ui.adsabs.harvard.edu/abs/1995AJ....109.1154F}
}

@ARTICLE{2022ApJ...940..170D,
       author = {{Dalessandro}, Emanuele and {Crociati}, Chiara and {Cignoni}, Michele and {Ferraro}, Francesco R. and {Lanzoni}, Barbara and {Origlia}, Livia and {Pallanca}, Cristina and {Rich}, R. Michael and {Saracino}, Sara and {Valenti}, Elena},
        title = "{Clues to the Formation of Liller 1 from Modeling Its Complex Star Formation History}",
      journal = {\apj},
         year = 2022,
        month = dec,
       volume = {940},
       number = {2},
          eid = {170},
        pages = {170},
          doi = {10.3847/1538-4357/ac9907},
archivePrefix = {arXiv},
       eprint = {2210.05694},
 primaryClass = {astro-ph.GA},
       adsurl = {https://ui.adsabs.harvard.edu/abs/2022ApJ...940..170D}
}

@ARTICLE{1977ApJ...213L..21L,
       author = {{Liller}, W.},
        title = "{Searches for the optical counterpart of the X-ray burst sources MXB 1728-34 and MXB 1730-33.}",
      journal = {\apjl},
         year = 1977,
        month = apr,
       volume = {213},
        pages = {L21-L23},
          doi = {10.1086/182401},
       adsurl = {https://ui.adsabs.harvard.edu/abs/1977ApJ...213L..21L}
}

@ARTICLE{2015ApJ...806..152S,
       author = {{Saracino}, S. and {Dalessandro}, E. and {Ferraro}, F.~R. and {Lanzoni}, B. and {Geisler}, D. and {Mauro}, F. and {Villanova}, S. and {Moni Bidin}, C. and {Miocchi}, P. and {Massari}, D.},
        title = "{GEMINI/GeMS Observations Unveil the Structure of the Heavily Obscured Globular Cluster Liller 1.}",
      journal = {\apj},
         year = 2015,
        month = jun,
       volume = {806},
       number = {2},
          eid = {152},
        pages = {152},
          doi = {10.1088/0004-637X/806/2/152},
archivePrefix = {arXiv},
       eprint = {1505.00568},
 primaryClass = {astro-ph.SR},
       adsurl = {https://ui.adsabs.harvard.edu/abs/2015ApJ...806..152S}
}

@ARTICLE{2001A&A...376..878O,
       author = {{Ortolani}, S. and {Barbuy}, B. and {Bica}, E. and {Renzini}, A. and {Zoccali}, M. and {Rich}, R.~M. and {Cassisi}, S.},
        title = "{HST NICMOS photometry of the reddened bulge globular clusters NGC 6528, Terzan 5, Liller 1, UKS 1 and Terzan 4}",
      journal = {\aap},
         year = 2001,
        month = sep,
       volume = {376},
        pages = {878-884},
          doi = {10.1051/0004-6361:20011045},
archivePrefix = {arXiv},
       eprint = {astro-ph/0107459},
 primaryClass = {astro-ph},
       adsurl = {https://ui.adsabs.harvard.edu/abs/2001A&A...376..878O}
}

@misc{bushouse_2025_17400413,
  author       = {Bushouse, Howard and
                  Eisenhamer, Jonathan and
                  Dencheva, Nadia and
                  Davies, James and
                  Greenfield, Perry and
                  Morrison, Jane and
                  Hodge, Phil and
                  Simon, Bernie and
                  Grumm, David and
                  Droettboom, Michael and
                  Slavich, Edward and
                  Sosey, Megan and
                  Pauly, Tyler and
                  Miller, Todd and
                  Jedrzejewski, Robert and
                  Hack, Warren and
                  Davis, David and
                  Crawford, Steven and
                  Law, David and
                  Gordon, Karl and
                  Regan, Michael and
                  Cara, Mihai and
                  MacDonald, Ken and
                  Bradley, Larry and
                  Shanahan, Clare and
                  Jamieson, William and
                  Teodoro, Mairan and
                  Williams, Thomas and
                  Pena-Guerrero, Maria and
                  Graham, Brett and
                  Molter, Edward and
                  Brandt, Timothy and
                  Hayes, Christian and
                  Cooper, Rachel and
                  Clarke, Melanie and
                  Filippazzo, Joseph},
  title        = {{JWST Calibration Pipeline}},
  month        = oct,
  year         = 2025,
  publisher    = {Zenodo},
  version      = {1.20.1},
  doi          = {10.5281/zenodo.17400413},
  url          = {https://doi.org/10.5281/zenodo.17400413}
}

@ARTICLE{2024MNRAS.533..756V,
       author = {{van den Eijnden}, J. and {Robins}, D. and {Sharma}, R. and {S{\'a}nchez-Fern{\'a}ndez}, C. and {Russell}, T.~D. and {Degenaar}, N. and {Miller-Jones}, J.~C.~A. and {Maccarone}, T.},
        title = "{The variable radio jet of the accreting neutron star the Rapid Burster}",
      journal = {\mnras},
         year = 2024,
        month = sep,
       volume = {533},
       number = {1},
        pages = {756-770},
          doi = {10.1093/mnras/stae1826},
archivePrefix = {arXiv},
       eprint = {2405.19827},
 primaryClass = {astro-ph.HE},
       adsurl = {https://ui.adsabs.harvard.edu/abs/2024MNRAS.533..756V}
}

@ARTICLE{2024ApJS..274...22E,
       author = {{Evans}, Ian N. and {Evans}, Janet D. and {Mart{\'\i}nez-Galarza}, J. Rafael and {Miller}, Joseph B. and {Primini}, Francis A. and {Azadi}, Mojegan and {Burke}, Douglas J. and {Civano}, Francesca M. and {D'Abrusco}, Raffaele and {Fabbiano}, Giuseppina and {Graessle}, Dale E. and {Grier}, John D. and {Houck}, John C. and {Lauer}, Jennifer and {McCollough}, Michael L. and {Nowak}, Michael A. and {Plummer}, David A. and {Rots}, Arnold H. and {Siemiginowska}, Aneta and {Tibbetts}, Michael S.},
        title = "{The Chandra Source Catalog Release 2 Series}",
      journal = {\apjs},
         year = 2024,
        month = oct,
       volume = {274},
       number = {2},
          eid = {22},
        pages = {22},
          doi = {10.3847/1538-4365/ad6319},
archivePrefix = {arXiv},
       eprint = {2407.10799},
 primaryClass = {astro-ph.HE},
       adsurl = {https://ui.adsabs.harvard.edu/abs/2024ApJS..274...22E}
}

@misc{2016ascl.soft08013D,
       author = {{Dolphin}, Andrew},
        title = "{DOLPHOT: Stellar photometry}",
 howpublished = {Astrophysics Source Code Library, record ascl:1608.013},
         year = 2016,
        month = aug,
          eid = {ascl:1608.013},
archivePrefix = {ascl},
       eprint = {1608.013},
       adsurl = {https://ui.adsabs.harvard.edu/abs/2016ascl.soft08013D}
}

@ARTICLE{2024ApJS..271...47W,
       author = {{Weisz}, Daniel R. and {Dolphin}, Andrew E. and {Savino}, Alessandro and {McQuinn}, Kristen B.~W. and {Newman}, Max J.~B. and {Williams}, Benjamin F. and {Kallivayalil}, Nitya and {Anderson}, Jay and {Boyer}, Martha L. and {Correnti}, Matteo and {Geha}, Marla C. and {Sandstrom}, Karin M. and {Cole}, Andrew A. and {Warfield}, Jack T. and {Skillman}, Evan D. and {Cohen}, Roger E. and {Beaton}, Rachael and {Bressan}, Alessandro and {Bolatto}, Alberto and {Boylan-Kolchin}, Michael and {Brooks}, Alyson M. and {Bullock}, James S. and {Conroy}, Charlie and {Cooper}, Michael C. and {Dalcanton}, Julianne J. and {Dotter}, Aaron L. and {Fritz}, Tobias K. and {Garling}, Christopher T. and {Gennaro}, Mario and {Gilbert}, Karoline M. and {Girardi}, Leo and {Johnson}, Benjamin D. and {Johnson}, L. Clifton and {Kalirai}, Jason and {Kirby}, Evan N. and {Lang}, Dustin and {Marigo}, Paola and {Richstein}, Hannah and {Schlafly}, Edward F. and {Tollerud}, Erik J. and {Wetzel}, Andrew},
        title = "{The JWST Resolved Stellar Populations Early Release Science Program. V. DOLPHOT Stellar Photometry for NIRCam and NIRISS}",
      journal = {\apjs},
         year = 2024,
        month = apr,
       volume = {271},
       number = {2},
          eid = {47},
        pages = {47},
          doi = {10.3847/1538-4365/ad2600},
archivePrefix = {arXiv},
       eprint = {2402.03504},
 primaryClass = {astro-ph.GA},
       adsurl = {https://ui.adsabs.harvard.edu/abs/2024ApJS..271...47W}
}

@misc{2022zndo...7487203B,
       author = {{Bushouse}, Howard and {Eisenhamer}, Jonathan and {Dencheva}, Nadia and {Davies}, James and {Greenfield}, Perry and {Morrison}, Jane and {Hodge}, Phil and {Simon}, Bernie and {Grumm}, David and {Droettboom}, Michael and {Slavich}, Edward and {Sosey}, Megan and {Pauly}, Tyler and {Miller}, Todd and {Jedrzejewski}, Robert and {Hack}, Warren and {Davis}, David and {Crawford}, Steven and {Law}, David and {Gordon}, Karl and {Regan}, Michael and {Cara}, Mihai and {MacDonald}, Ken and {Bradley}, Larry and {Shanahan}, Clare and {Jamieson}, William and {Teodoro}, Mairan and {Williams}, Thomas},
        title = "{JWST Calibration Pipeline}",
         year = 2022,
        month = dec,
          eid = {10.5281/zenodo.7487203},
          doi = {10.5281/zenodo.7487203},
      version = {1.9.0},
    publisher = {Zenodo},
       adsurl = {https://ui.adsabs.harvard.edu/abs/2022zndo...7487203B}
}

@INPROCEEDINGS{2012SPIE.8442E..3DP,
       author = {{Perrin}, Marshall D. and {Soummer}, R{\'e}mi and {Elliott}, Erin M. and {Lallo}, Matthew D. and {Sivaramakrishnan}, Anand},
        title = "{Simulating point spread functions for the James Webb Space Telescope with WebbPSF}",
    booktitle = {Space Telescopes and Instrumentation 2012: Optical, Infrared, and Millimeter Wave},
         year = 2012,
       editor = {{Clampin}, Mark C. and {Fazio}, Giovanni G. and {MacEwen}, Howard A. and {Oschmann}, Jr., Jacobus M.},
       series = {Society of Photo-Optical Instrumentation Engineers (SPIE) Conference Series},
       volume = {8442},
        month = sep,
          eid = {84423D},
        pages = {84423D},
          doi = {10.1117/12.925230},
       adsurl = {https://ui.adsabs.harvard.edu/abs/2012SPIE.8442E..3DP}
}

@ARTICLE{2021MNRAS.505.5978V,
       author = {{Vasiliev}, Eugene and {Baumgardt}, Holger},
        title = "{Gaia EDR3 view on galactic globular clusters}",
      journal = {\mnras},
         year = 2021,
        month = aug,
       volume = {505},
       number = {4},
        pages = {5978-6002},
          doi = {10.1093/mnras/stab1475},
archivePrefix = {arXiv},
       eprint = {2102.09568},
 primaryClass = {astro-ph.GA},
       adsurl = {https://ui.adsabs.harvard.edu/abs/2021MNRAS.505.5978V}
}

@ARTICLE{2024ApJ...969L...7Y,
       author = {{Yin}, Dejiang and {Zhang}, Li-yun and {Qian}, Lei and {Eatough}, Ralph P. and {Li}, Baoda and {Lorimer}, Duncan R. and {Dai}, Yinfeng and {Li}, Yaowei and {Zhang}, Xingnan and {Li}, Minghui and {Su}, Tianhao and {Wu}, Yuxiao and {Pan}, Yu and {Lian}, Yujie and {Liu}, Tong and {Yan}, Zhen and {Pan}, Zhichen},
        title = "{FAST Discovery of Eight Isolated Millisecond Pulsars in NGC 6517}",
      journal = {\apjl},
         year = 2024,
        month = jul,
       volume = {969},
       number = {1},
          eid = {L7},
        pages = {L7},
          doi = {10.3847/2041-8213/ad534e},
archivePrefix = {arXiv},
       eprint = {2405.18228},
 primaryClass = {astro-ph.HE},
       adsurl = {https://ui.adsabs.harvard.edu/abs/2024ApJ...969L...7Y}
}

@ARTICLE{2011ApJ...729...90T,
       author = {{Tam}, P.~H.~T. and {Kong}, A.~K.~H. and {Hui}, C.~Y. and {Cheng}, K.~S. and {Li}, C. and {Lu}, T.-N.},
        title = "{Gamma-ray Emission from the Globular Clusters Liller 1, M80, NGC 6139, NGC 6541, NGC 6624, and NGC 6752}",
      journal = {\apj},
         year = 2011,
        month = mar,
       volume = {729},
       number = {2},
          eid = {90},
        pages = {90},
          doi = {10.1088/0004-637X/729/2/90},
archivePrefix = {arXiv},
       eprint = {1101.4106},
 primaryClass = {astro-ph.HE},
       adsurl = {https://ui.adsabs.harvard.edu/abs/2011ApJ...729...90T}
}

@ARTICLE{KBurdgeinprep,
       author = {{Burdge}, K.~B. and {Caiazzo}, I. and {Desai}, M.~M. and {Pallanca}, C. and {Berger}, V.~L. and {Brauer}, K. and {Correnti}, M. and {Draghis}, P. and {El-Badry}, K. and {Ferraro}, F.~R. and {Gonz´alez-Caniulef}, D. and {Kremer}, K. and {Lanzoni}, B. and {Li}, D. and {Lu}, J.~R. and {Mo}, G. and {Ng}, M. and {Romani}, R.~W. and {Shin}, K. and {Weisz}, D.~R. and {Zullo}, G.},
        title = "{Discovery of over a thousand variable stars in Terzan 5 and Liller 1 with JWST}",
      journal = {\apj, submitted},
         year = 2026
}

@ARTICLE{2009PASJ...61..999M,
       author = {{Matsuoka}, Masaru and {Kawasaki}, Kazuyoshi and {Ueno}, Shiro and {Tomida}, Hiroshi and {Kohama}, Mitsuhiro and {Suzuki}, Motoko and {Adachi}, Yasuki and {Ishikawa}, Masaki and {Mihara}, Tatehiro and {Sugizaki}, Mutsumi and {Isobe}, Naoki and {Nakagawa}, Yujin and {Tsunemi}, Hiroshi and {Miyata}, Emi and {Kawai}, Nobuyuki and {Kataoka}, Jun and {Morii}, Mikio and {Yoshida}, Atsumasa and {Negoro}, Hitoshi and {Nakajima}, Motoki and {Ueda}, Yoshihiro and {Chujo}, Hirotaka and {Yamaoka}, Kazutaka and {Yamazaki}, Osamu and {Nakahira}, Satoshi and {You}, Tetsuya and {Ishiwata}, Ryoji and {Miyoshi}, Sho and {Eguchi}, Satoshi and {Hiroi}, Kazuo and {Katayama}, Haruyoshi and {Ebisawa}, Ken},
        title = "{The MAXI Mission on the ISS: Science and Instruments for Monitoring All-Sky X-Ray Images}",
      journal = {\pasj},
         year = 2009,
        month = oct,
       volume = {61},
        pages = {999},
          doi = {10.1093/pasj/61.5.999},
archivePrefix = {arXiv},
       eprint = {0906.0631},
 primaryClass = {astro-ph.IM},
       adsurl = {https://ui.adsabs.harvard.edu/abs/2009PASJ...61..999M}
}

@ARTICLE{2025ApJS..279...57H,
       author = {{Heinke}, Craig O. and {Zheng}, Junwen and {Maccarone}, Thomas J. and {Degenaar}, Nathalie and {Bahramian}, Arash and {Sivakoff}, Gregory R. and {Toor}, Simrat},
        title = "{Catalog of Outbursts of Neutron Star Low-mass X-Ray Binaries}",
      journal = {\apjs},
         year = 2025,
        month = aug,
       volume = {279},
       number = {2},
          eid = {57},
        pages = {57},
          doi = {10.3847/1538-4365/ade99a},
archivePrefix = {arXiv},
       eprint = {2407.18867},
 primaryClass = {astro-ph.HE},
       adsurl = {https://ui.adsabs.harvard.edu/abs/2025ApJS..279...57H}
}

@ARTICLE{1977Natur.270..310J,
       author = {{Joss}, P.~C.},
        title = "{X-ray bursts and neutron-star thermonuclear flashes}",
      journal = {\nat},
         year = 1977,
        month = nov,
       volume = {270},
       number = {5635},
        pages = {310-314},
          doi = {10.1038/270310a0},
       adsurl = {https://ui.adsabs.harvard.edu/abs/1977Natur.270..310J}
}

@ARTICLE{2007ApJ...661..468C,
       author = {{Cooper}, Randall L. and {Narayan}, Ramesh},
        title = "{Hydrogen-triggered Type I X-Ray Bursts in a Two-Zone Model}",
      journal = {\apj},
         year = 2007,
        month = may,
       volume = {661},
       number = {1},
        pages = {468-476},
          doi = {10.1086/513461},
archivePrefix = {arXiv},
       eprint = {astro-ph/0702042},
 primaryClass = {astro-ph},
       adsurl = {https://ui.adsabs.harvard.edu/abs/2007ApJ...661..468C}
}

@ARTICLE{1996ApJ...462L..39L,
       author = {{Lewin}, Walter H.~G. and {Rutledge}, Robert E. and {Kommers}, Jefferson M. and {van Paradijs}, Jan and {Kouveliotou}, Chryssa},
        title = "{A Comparison between the Rapid Burster and GRO J1744-28}",
      journal = {\apjl},
         year = 1996,
        month = may,
       volume = {462},
        pages = {L39},
          doi = {10.1086/310022},
       adsurl = {https://ui.adsabs.harvard.edu/abs/1996ApJ...462L..39L}
}

@ARTICLE{1996ApJ...469L..29Z,
       author = {{Zhang}, W. and {Morgan}, E.~H. and {Jahoda}, K. and {Swank}, J.~H. and {Strohmayer}, T.~E. and {Jernigan}, G. and {Klein}, R.~I.},
        title = "{Quasi-periodic X-Ray Brightness Oscillations of GRO J1744-28}",
      journal = {\apjl},
         year = 1996,
        month = sep,
       volume = {469},
        pages = {L29},
          doi = {10.1086/310259},
       adsurl = {https://ui.adsabs.harvard.edu/abs/1996ApJ...469L..29Z}
}

@ARTICLE{1996IAUC.6286....1K,
       author = {{Kouveliotou}, C. and {Kommers}, J. and {Lewin}, W.~H.~G. and {van Paradijs}, J. and {Fishman}, G.~J. and {Briggs}, M.~S. and {Hurley}, K. and {Harmon}, A. and {Finger}, M.~H. and {Wilson}, R.~B.},
        title = "{GRO J1744-28}",
      journal = {\iaucirc},
         year = 1996,
        month = jan,
       volume = {6286},
        pages = {1},
       adsurl = {https://ui.adsabs.harvard.edu/abs/1996IAUC.6286....1K}
}

@ARTICLE{1997ApJ...486..435R,
       author = {{Rappaport}, S. and {Joss}, P.~C.},
        title = "{The Nature and Evolutionary History of GRO J1744-28}",
      journal = {\apj},
         year = 1997,
        month = sep,
       volume = {486},
       number = {1},
        pages = {435-444},
          doi = {10.1086/304506},
       adsurl = {https://ui.adsabs.harvard.edu/abs/1997ApJ...486..435R}
}

@misc{intZand2025bursterlist,
  author       = {in 't Zand, Jean J. M.},
  title        = {List of 122 {Galactic} {Type-I} {X-ray} Bursters},
  howpublished = {\url{https://sronpersonalpages.nl/~jeanz/bursterlist.html}},
  note         = {Updated 10 October 2025},
  year         = {2025},
  urldate      = {2026-07-28}
}

@ARTICLE{1997A&A...321L..25L,
       author = {{Li}, X.-D. and {van den Heuvel}, E.~P.~J.},
        title = "{On the nature of SMC X-1.}",
      journal = {\aap},
         year = 1997,
        month = may,
       volume = {321},
        pages = {L25-L28},
       adsurl = {https://ui.adsabs.harvard.edu/abs/1997A&A...321L..25L}
}

@ARTICLE{2018RAA....18..148R,
       author = {{Rai}, Binay and {Pradhan}, Pragati and {Paul}, Bikash Chandra},
        title = "{A report on Type II X-ray bursts from SMC X-1}",
      journal = {Research in Astronomy and Astrophysics},
         year = 2018,
        month = dec,
       volume = {18},
       number = {12},
          eid = {148},
        pages = {148},
          doi = {10.1088/1674-4527/18/12/148},
archivePrefix = {arXiv},
       eprint = {1806.03244},
 primaryClass = {astro-ph.HE},
       adsurl = {https://ui.adsabs.harvard.edu/abs/2018RAA....18..148R}
}

@ARTICLE{2020ApJ...895...10P,
       author = {{Pradhan}, Pragati and {Maitra}, Chandreyee and {Paul}, Biswajit},
        title = "{Is Superorbital Modulation in SMC X-1 Caused by Absorption in a Warped Precessing Accretion Disk?}",
      journal = {\apj},
         year = 2020,
        month = may,
       volume = {895},
       number = {1},
          eid = {10},
        pages = {10},
          doi = {10.3847/1538-4357/ab8224},
archivePrefix = {arXiv},
       eprint = {2004.00664},
 primaryClass = {astro-ph.HE},
       adsurl = {https://ui.adsabs.harvard.edu/abs/2020ApJ...895...10P}
}

@ARTICLE{1978ApJ...225L.123J,
       author = {{Joss}, P.~C.},
        title = "{Helium-burning flashes on an accreting neutron star: a model for X-ray burst sources.}",
      journal = {\apjl},
         year = 1978,
        month = nov,
       volume = {225},
        pages = {L123-L127},
          doi = {10.1086/182808},
       adsurl = {https://ui.adsabs.harvard.edu/abs/1978ApJ...225L.123J}
}

@ARTICLE{2003ApJ...582L..91M,
       author = {{Moon}, Dae-Sik and {Eikenberry}, Stephen S. and {Wasserman}, Ira M.},
        title = "{SMC X-1 as an Intermediate-Stage Flaring X-Ray Pulsar}",
      journal = {\apjl},
         year = 2003,
        month = jan,
       volume = {582},
       number = {2},
        pages = {L91-L94},
          doi = {10.1086/367782},
archivePrefix = {arXiv},
       eprint = {astro-ph/0209414},
 primaryClass = {astro-ph},
       adsurl = {https://ui.adsabs.harvard.edu/abs/2003ApJ...582L..91M}
}

@ARTICLE{2002A&A...381L..45M,
       author = {{Masetti}, N.},
        title = "{A change in the outburst recurrence time of the Rapid Burster}",
      journal = {\aap},
         year = 2002,
        month = jan,
       volume = {381},
        pages = {L45-L48},
          doi = {10.1051/0004-6361:20011642},
archivePrefix = {arXiv},
       eprint = {astro-ph/0111382},
 primaryClass = {astro-ph},
       adsurl = {https://ui.adsabs.harvard.edu/abs/2002A&A...381L..45M}
}

@ARTICLE{2007A&G....48e..12M,
       author = {{Maccarone}, Tom and {Knigge}, Christian},
        title = "{Compact objects in globular clusters}",
      journal = {Astronomy and Geophysics},
         year = 2007,
        month = oct,
       volume = {48},
       number = {5},
        pages = {5.12-5.20},
          doi = {10.1111/j.1468-4004.2007.48512.x},
archivePrefix = {arXiv},
       eprint = {0709.3732},
 primaryClass = {astro-ph},
       adsurl = {https://ui.adsabs.harvard.edu/abs/2007A&G....48e..12M}
}

@INPROCEEDINGS{2026enap....3..458K,
       author = {{Kremer}, Kyle},
        title = "{Compact objects in globular clusters}",
    booktitle = {Encyclopedia of Astrophysics, Volume 3},
         year = 2026,
       volume = {3},
        month = jan,
        pages = {458-472},
          doi = {10.1016/B978-0-443-21439-4.00103-6},
archivePrefix = {arXiv},
       eprint = {2508.14308},
 primaryClass = {astro-ph.HE},
       adsurl = {https://ui.adsabs.harvard.edu/abs/2026enap....3..458K}
}

@misc{2016ascl.soft09011B,
       author = {{Bradley}, Larry and {Sipocz}, Brigitta and {Robitaille}, Thomas and {Tollerud}, Erik and {Deil}, Christoph and {Vin{\'\i}cius}, Z{\`e} and {Barbary}, Kyle and {G{\"u}nther}, Hans Moritz and {Bostroem}, Azalee and {Droettboom}, Michael and {Bray}, Erik and {Bratholm}, Lars Andersen and {Pickering}, T.~E. and {Craig}, Matt and {Pascual}, Sergio and {Greco}, Johnny and {Donath}, Axel and {Kerzendorf}, Wolfgang and {Littlefair}, Stuart and {Barentsen}, Geert and {D'Eugenio}, Francesco and {Weaver}, Benjamin Alan},
        title = "{Photutils: Photometry tools}",
 howpublished = {Astrophysics Source Code Library, record ascl:1609.011},
         year = 2016,
        month = sep,
          eid = {ascl:1609.011},
archivePrefix = {ascl},
       eprint = {1609.011},
       adsurl = {https://ui.adsabs.harvard.edu/abs/2016ascl.soft09011B}
}

@ARTICLE{1987PASP...99..191S,
       author = {{Stetson}, Peter B.},
        title = "{DAOPHOT: A Computer Program for Crowded-Field Stellar Photometry}",
      journal = {\pasp},
         year = 1987,
        month = mar,
       volume = {99},
        pages = {191},
          doi = {10.1086/131977},
       adsurl = {https://ui.adsabs.harvard.edu/abs/1987PASP...99..191S}
}

@ARTICLE{1982Natur.300..728A,
       author = {{Alpar}, M.~A. and {Cheng}, A.~F. and {Ruderman}, M.~A. and {Shaham}, J.},
        title = "{A new class of radio pulsars}",
      journal = {\nat},
         year = 1982,
        month = dec,
       volume = {300},
       number = {5894},
        pages = {728-730},
          doi = {10.1038/300728a0},
       adsurl = {https://ui.adsabs.harvard.edu/abs/1982Natur.300..728A}
}

@ARTICLE{2023A&A...674A...1G,
       author = {{Gaia Collaboration} and {Vallenari}, A. and {Brown}, A.~G.~A. and {Prusti}, T. and {de Bruijne}, J.~H.~J. and {Arenou}, F. and {Babusiaux}, C. and {Biermann}, M. and {Creevey}, O.~L. and {Ducourant}, C. and {Evans}, D.~W. and {Eyer}, L. and {Guerra}, R. and {Hutton}, A. and {Jordi}, C. and {Klioner}, S.~A. and {Lammers}, U.~L. and {Lindegren}, L. and {Luri}, X. and {Mignard}, F. and {Panem}, C. and {Pourbaix}, D. and {Randich}, S. and {Sartoretti}, P. and {Soubiran}, C. and {Tanga}, P. and {Walton}, N.~A. and {Bailer-Jones}, C.~A.~L. and {Bastian}, U. and {Drimmel}, R. and {Jansen}, F. and {Katz}, D. and {Lattanzi}, M.~G. and {van Leeuwen}, F. and {Bakker}, J. and {Cacciari}, C. and {Casta{\~n}eda}, J. and {De Angeli}, F. and {Fabricius}, C. and {Fouesneau}, M. and {Fr{\'e}mat}, Y. and {Galluccio}, L. and {Guerrier}, A. and {Heiter}, U. and {Masana}, E. and {Messineo}, R. and {Mowlavi}, N. and {Nicolas}, C. and {Nienartowicz}, K. and {Pailler}, F. and {Panuzzo}, P. and {Riclet}, F. and {Roux}, W. and {Seabroke}, G.~M. and {Sordo}, R. and {Th{\'e}venin}, F. and {Gracia-Abril}, G. and {Portell}, J. and {Teyssier}, D. and {Altmann}, M. and {Andrae}, R. and {Audard}, M. and {Bellas-Velidis}, I. and {Benson}, K. and {Berthier}, J. and {Blomme}, R. and {Burgess}, P.~W. and {Busonero}, D. and {Busso}, G. and {C{\'a}novas}, H. and {Carry}, B. and {Cellino}, A. and {Cheek}, N. and {Clementini}, G. and {Damerdji}, Y. and {Davidson}, M. and {de Teodoro}, P. and {Nu{\~n}ez Campos}, M. and {Delchambre}, L. and {Dell'Oro}, A. and {Esquej}, P. and {Fern{\'a}ndez-Hern{\'a}ndez}, J. and {Fraile}, E. and {Garabato}, D. and {Garc{\'\i}a-Lario}, P. and {Gosset}, E. and {Haigron}, R. and {Halbwachs}, J.-L. and {Hambly}, N.~C. and {Harrison}, D.~L. and {Hern{\'a}ndez}, J. and {Hestroffer}, D. and {Hodgkin}, S.~T. and {Holl}, B. and {Jan{\ss}en}, K. and {Jevardat de Fombelle}, G. and {Jordan}, S. and {Krone-Martins}, A. and {Lanzafame}, A.~C. and {L{\"o}ffler}, W. and {Marchal}, O. and {Marrese}, P.~M. and {Moitinho}, A. and {Muinonen}, K. and {Osborne}, P. and {Pancino}, E. and {Pauwels}, T. and {Recio-Blanco}, A. and {Reyl{\'e}}, C. and {Riello}, M. and {Rimoldini}, L. and {Roegiers}, T. and {Rybizki}, J. and {Sarro}, L.~M. and {Siopis}, C. and {Smith}, M. and {Sozzetti}, A. and {Utrilla}, E. and {van Leeuwen}, M. and {Abbas}, U. and {{\'A}brah{\'a}m}, P. and {Abreu Aramburu}, A. and {Aerts}, C. and {Aguado}, J.~J. and {Ajaj}, M. and {Aldea-Montero}, F. and {Altavilla}, G. and {{\'A}lvarez}, M.~A. and {Alves}, J. and {Anders}, F. and {Anderson}, R.~I. and {Anglada Varela}, E. and {Antoja}, T. and {Baines}, D. and {Baker}, S.~G. and {Balaguer-N{\'u}{\~n}ez}, L. and {Balbinot}, E. and {Balog}, Z. and {Barache}, C. and {Barbato}, D. and {Barros}, M. and {Barstow}, M.~A. and {Bartolom{\'e}}, S. and {Bassilana}, J.-L. and {Bauchet}, N. and {Becciani}, U. and {Bellazzini}, M. and {Berihuete}, A. and {Bernet}, M. and {Bertone}, S. and {Bianchi}, L. and {Binnenfeld}, A. and {Blanco-Cuaresma}, S. and {Blazere}, A. and {Boch}, T. and {Bombrun}, A. and {Bossini}, D. and {Bouquillon}, S. and {Bragaglia}, A. and {Bramante}, L. and {Breedt}, E. and {Bressan}, A. and {Brouillet}, N. and {Brugaletta}, E. and {Bucciarelli}, B. and {Burlacu}, A. and {Butkevich}, A.~G. and {Buzzi}, R. and {Caffau}, E. and {Cancelliere}, R. and {Cantat-Gaudin}, T. and {Carballo}, R. and {Carlucci}, T. and {Carnerero}, M.~I. and {Carrasco}, J.~M. and {Casamiquela}, L. and {Castellani}, M. and {Castro-Ginard}, A. and {Chaoul}, L. and {Charlot}, P. and {Chemin}, L. and {Chiaramida}, V. and {Chiavassa}, A. and {Chornay}, N. and {Comoretto}, G. and {Contursi}, G. and {Cooper}, W.~J. and {Cornez}, T. and {Cowell}, S. and {Crifo}, F. and {Cropper}, M. and {Crosta}, M. and {Crowley}, C. and {Dafonte}, C. and {Dapergolas}, A. and {David}, M. and {David}, P. and {de Laverny}, P. and {De Luise}, F. and {De March}, R.},
        title = "{Gaia Data Release 3. Summary of the content and survey properties}",
      journal = {\aap},
         year = 2023,
        month = jun,
       volume = {674},
          eid = {A1},
        pages = {A1},
          doi = {10.1051/0004-6361/202243940},
archivePrefix = {arXiv},
       eprint = {2208.00211},
 primaryClass = {astro-ph.GA},
       adsurl = {https://ui.adsabs.harvard.edu/abs/2023A&A...674A...1G}
}

@ARTICLE{1989ApJ...345..245C,
       author = {{Cardelli}, Jason A. and {Clayton}, Geoffrey C. and {Mathis}, John S.},
        title = "{The Relationship between Infrared, Optical, and Ultraviolet Extinction}",
      journal = {\apj},
         year = 1989,
        month = oct,
       volume = {345},
        pages = {245},
          doi = {10.1086/167900},
       adsurl = {https://ui.adsabs.harvard.edu/abs/1989ApJ...345..245C}
}

@ARTICLE{2017arXiv170307221I,
       author = {{in 't Zand}, J.~J.~M. and {Bagnoli}, T. and {D'Angelo}, C. and {Patruno}, A. and {Galloway}, D.~K. and {van der Klis}, M.~B.~M. and {Watts}, A.~L. and {Marshall}, H.~L.},
        title = "{Chandra spectroscopy of Rapid Burster type-I X-ray bursts}",
      journal = {arXiv e-prints},
         year = 2017,
        month = mar,
          eid = {arXiv:1703.07221},
        pages = {arXiv:1703.07221},
          doi = {10.48550/arXiv.1703.07221},
archivePrefix = {arXiv},
       eprint = {1703.07221},
 primaryClass = {astro-ph.HE},
       adsurl = {https://ui.adsabs.harvard.edu/abs/2017arXiv170307221I}
}

@ARTICLE{2012ApJ...752..158S,
       author = {{Sala}, G. and {Haberl}, F. and {Jos{\'e}}, J. and {Parikh}, A. and {Longland}, R. and {Pardo}, L.~C. and {Andersen}, M.},
        title = "{Constraints on the Mass and Radius of the Accreting Neutron Star in the Rapid Burster}",
      journal = {\apj},
         year = 2012,
        month = jun,
       volume = {752},
       number = {2},
          eid = {158},
        pages = {158},
          doi = {10.1088/0004-637X/752/2/158},
archivePrefix = {arXiv},
       eprint = {1204.3627},
 primaryClass = {astro-ph.HE},
       adsurl = {https://ui.adsabs.harvard.edu/abs/2012ApJ...752..158S}
}

@ARTICLE{1999PASP..111...63F,
       author = {{Fitzpatrick}, Edward L.},
        title = "{Correcting for the Effects of Interstellar Extinction}",
      journal = {\pasp},
         year = 1999,
        month = jan,
       volume = {111},
       number = {755},
        pages = {63-75},
          doi = {10.1086/316293},
archivePrefix = {arXiv},
       eprint = {astro-ph/9809387},
 primaryClass = {astro-ph},
       adsurl = {https://ui.adsabs.harvard.edu/abs/1999PASP..111...63F}
}

@ARTICLE{2019MNRAS.482.5138B,
       author = {{Baumgardt}, H. and {Hilker}, M. and {Sollima}, A. and {Bellini}, A.},
        title = "{Mean proper motions, space orbits, and velocity dispersion profiles of Galactic globular clusters derived from Gaia DR2 data}",
      journal = {\mnras},
         year = 2019,
        month = feb,
       volume = {482},
       number = {4},
        pages = {5138-5155},
          doi = {10.1093/mnras/sty2997},
archivePrefix = {arXiv},
       eprint = {1811.01507},
 primaryClass = {astro-ph.GA},
       adsurl = {https://ui.adsabs.harvard.edu/abs/2019MNRAS.482.5138B}
}

@ARTICLE{2020MNRAS.495L..37V,
       author = {{Vincentelli}, F.~M. and {Cavecchi}, Y. and {Casella}, P. and {Migliari}, S. and {Altamirano}, D. and {Belloni}, T. and {Diaz-Trigo}, M.},
        title = "{Discovery of a thermonuclear Type I X-ray burst in infrared: new limits on the orbital period of 4U 1728-34}",
      journal = {\mnras},
         year = 2020,
        month = jun,
       volume = {495},
       number = {1},
        pages = {L37-L41},
          doi = {10.1093/mnrasl/slaa049},
archivePrefix = {arXiv},
       eprint = {2003.08403},
 primaryClass = {astro-ph.HE},
       adsurl = {https://ui.adsabs.harvard.edu/abs/2020MNRAS.495L..37V}
}
\bibliographystyle{aasjournalv7}

\end{document}